\documentclass[prl,showpacs,amsmath,amssymb,floatfix]{revtex4}
\usepackage[latin9]{inputenc}
\usepackage{color}
\usepackage[caption=false]{subfig}
\usepackage{verbatim}
\usepackage{graphicx}
\usepackage{esint}
\usepackage{epstopdf}

\makeatletter
\@ifundefined{textcolor}{}
{%
 \definecolor{BLACK}{gray}{0}
 \definecolor{WHITE}{gray}{1}
 \definecolor{RED}{rgb}{1,0,0}
 \definecolor{GREEN}{rgb}{0,1,0}
 \definecolor{BLUE}{rgb}{0,0,1}
 \definecolor{CYAN}{cmyk}{1,0,0,0}
 \definecolor{MAGENTA}{cmyk}{0,1,0,0}
 \definecolor{YELLOW}{cmyk}{0,0,1,0}
}

\usepackage{epsfig}
\usepackage{dcolumn}
\usepackage{bm}
\def\(({\left(}
\def\)){\right)}
\def\[[{\left[}
\def\]]{\right]}

\newcommand{\be}{\begin{equation}}
\newcommand{\ee}{\end{equation}}
\newcommand{\bea}{\begin{eqnarray}}
\newcommand{\eea}{\end{eqnarray}}

\makeatother

\begin{document}

\title{Extending the applicability of Thermal Dynamics to Evolutionary Biology.}

\author{Tommaso Brotto$^{1,2}$, Guy Bunin$^{3}$ and Jorge Kurchan$^{1}$}

\affiliation{
$^{1}$Laboratoire de Physique Statistique de l'Ecole Normale Sup\'erieure,
CNRS UMR 8550 - Universit\'e Paris 6 - Universit\'e Paris 7;
24, rue Lhomond, 75005 Paris, France
$^{2}$ Dipartimento di Fisica, Universit\`a degli Studi di Milano,
Via Celoria 16, 20133 Milano, Italy. INFN, Sezione di Milano, Via Celoria 16, 20133 Milano, Italy\\
$^{3}$ Massachusetts Institute of Technology, Department of Physics,
Cambridge, Massachusetts 02139, USA}

\begin{abstract}
In the past years, a remarkable mapping has been found between the
dynamics of a population of $M$ individuals undergoing random mutations
and selection, and that of a single system in contact with a thermal
bath with temperature $1/M$. This correspondence holds under the
somewhat restrictive condition that the population is dominated by
a single type at almost all times, punctuated by rare successive mutations.
Here we argue that such thermal dynamics will hold more generally,
specifically in systems with rugged fitness landscapes. This includes
cases with strong clonal interference, where a number of concurrent
mutants dominate the population. The problem becomes closely analogous
to the experimental situation of glasses subjected to controlled variations
of parameters such as temperature, pressure or magnetic fields.
Non-trivial suggestions from the  field of glasses may be thus proposed for evolutionary
systems -- including a large part of the numerical simulation procedures -- that in many cases
would have been counter intuitive without this background. 
\end{abstract}
\maketitle
\vspace{0.5cm}

\section{Introduction}

\vspace{0.5cm}

\begin{figure}
\centering 
\subfloat[]{\includegraphics[angle=-90,width=0.32\columnwidth]{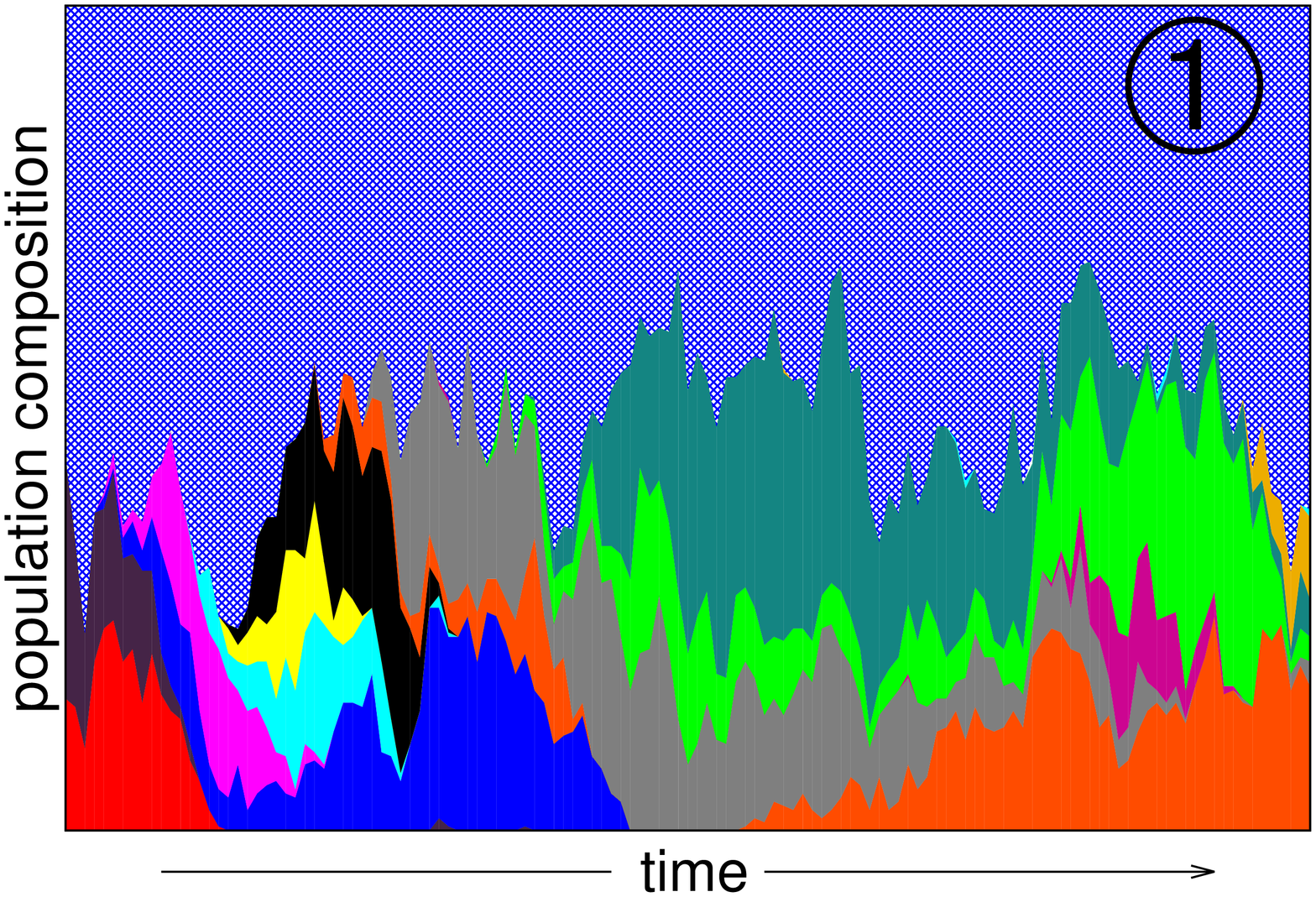}\label{concu1a}}
\subfloat[]{\includegraphics[angle=-90,width=0.32\columnwidth]{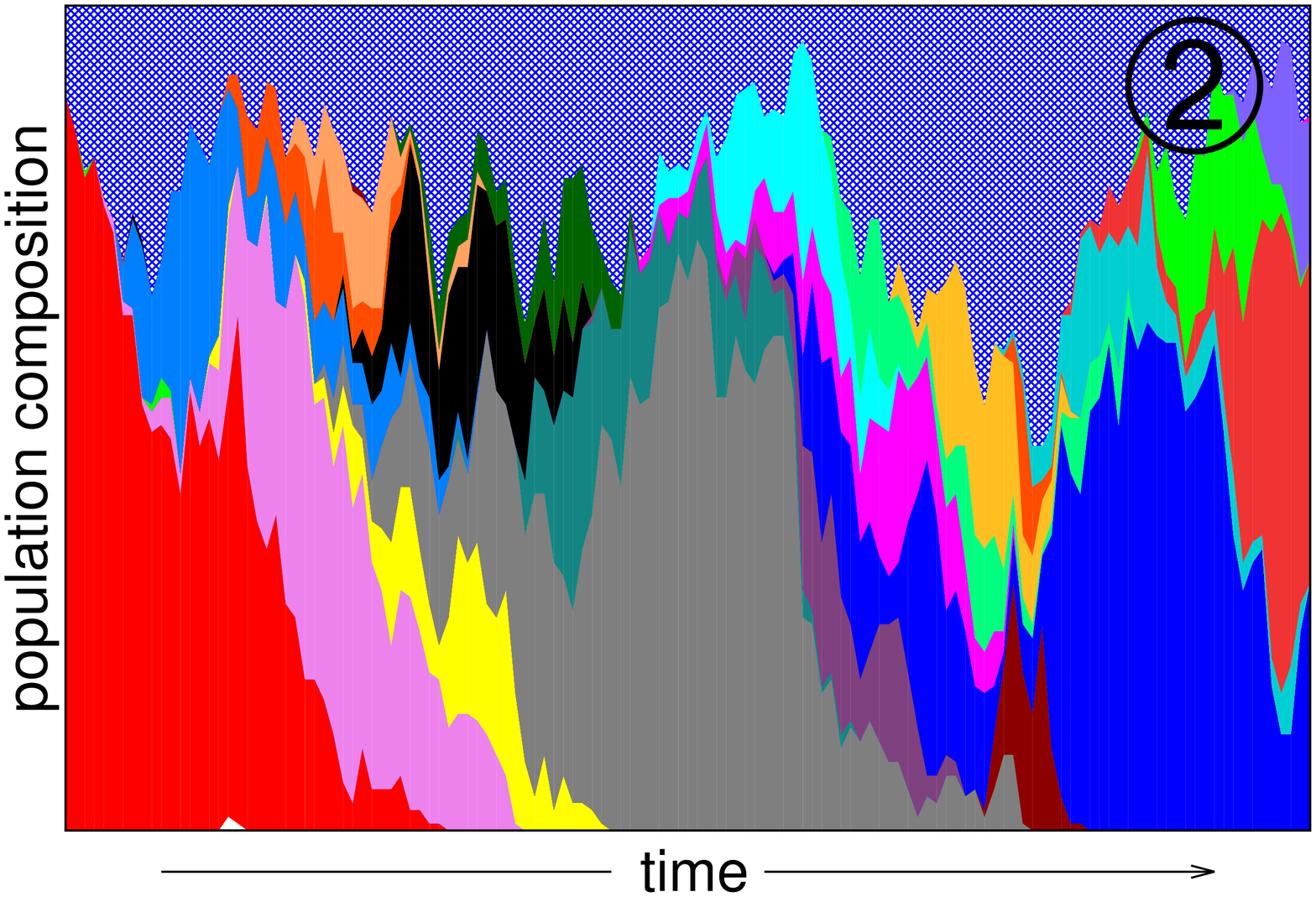}\label{concu1b}}
\subfloat[]{\includegraphics[angle=-90,width=0.32\columnwidth]{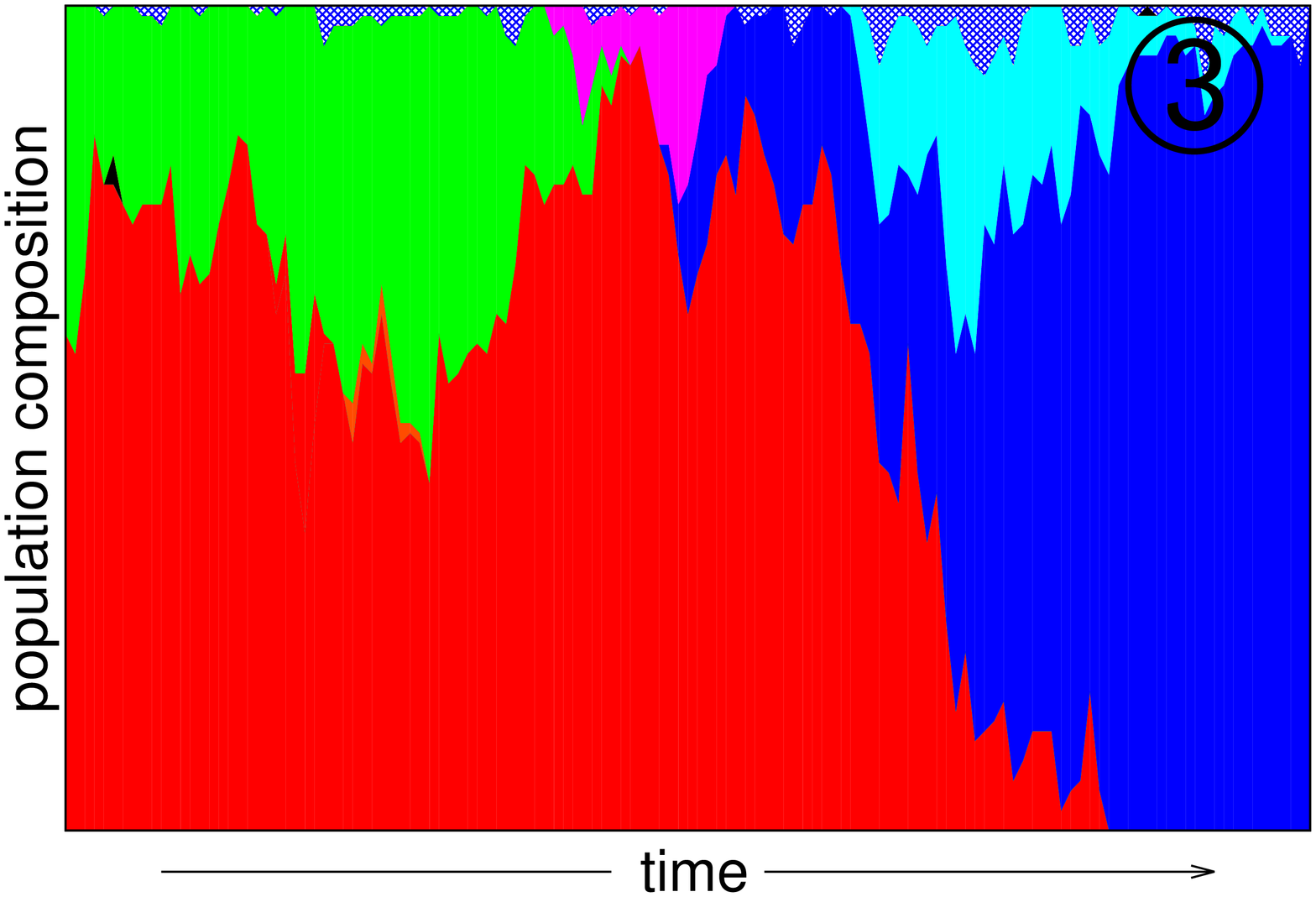}\label{concu1c}}
\protect\protect\protect\protect\caption{Mutant frequencies vs. time for the model in section \ref{XorSatExample}. Different
colors identify different mutants, and blue mesh includes all types
that never reach 10\% of the total population size. In (c), the population
at almost all times is dominated by a single mutant, whose identity
is replaced on rapid, successive sweeps. Detailed-balance is known
to hold here. In (a+b) more complex patterns are observed. Here too,
as we show below (section \ref{XorSatExample}), detailed-balance holds, provided that
proper averaging on short-times is applied. The labels of the figures
help locating them on the phase diagram of Fig \ref{phasediagram}.}
\label{concu1} 
\end{figure}

The dynamics of a population of $M$ individuals reproducing and undergoing
random mutations and selection has long been recognized to bear a
resemblance with a system driven by a `fitness potential', with an element
of `noise' given by random fluctuations that are the larger, the smaller
the total population (see, e.g. \cite{crow_introduction_1970,peliti_introduction_1997}).
However, the stochastic dynamics of a system in contact with a thermal
bath satisfy the relation of `detailed-balance' -- the condition that
the bath is itself in thermal equilibrium -- obviously not applicable
in general to an evolutionary dynamics with mutation and selection.
A known exception happens when the population is dominated by a single
mutant at any time, whose identity changes in rare and rapid `sweeps'
in which a new mutant fixes \cite{tsimring_rna_1996,lynch_lower_2011,neher_emergence_2013,sella_application_2005,mustonen_fitness_2010},
see Fig. \ref{concu1c} \cite{noteone}. It turns out that in that
special case \cite{berg_stochastic_2003,berg_adaptive_2004,sella_application_2005,barton_application_2009},
the analogy becomes a strict mathematical correspondence, opening
the door to statistical mechanical reasoning for evolving systems.

In spite of its potential, this remarkable and somewhat counter-intuitive
mapping has attracted relatively modest attention among physicists,
partly because of the restrictive conditions necessary for it to hold.
The purpose of this paper is to show that the correspondence to thermal
dynamics is in fact, given an appropriate interpretation of observations,
more robust for rugged fitness landscapes, where the changes in fitness
in each beneficial mutation are small. Importantly, the same temperature
is associated with the thermal dynamics in all cases.

As we shall see, the basic idea is that an essential ingredient for
the correspondence is a timescale-separation between the rate of reproduction
and the rate of fitness improvement, a separation which arises naturally
in the evolution on rugged landscapes (see e.g. \cite{elena_microbial_2003})
We then argue that {\em provided that a proper averaging of observables
over short times is applied}, one may extend the applicability of
detailed-balance, including conditions where \textcolor{black}{several different mutant lineages compete 
simultaneously for fixation} (clonal interference), and the changes
in fitness in a single mutation (selective advantage) are less restrictive
than before.

This paper is organized as follows: in section \ref{EmergThermal} we review the emergence
of thermal dynamics, using the `House of Cards' model as an example.
We also discuss there the limitations of the development. In section \ref{TimescaleSep} 
we very briefly review an extremely general, and in this context
very important fact: a system that is optimizing slowly will necessarily
exhibit fast and slow degrees of freedom, the timescale-separation
between these increases as the system evolves, and, concomitantly, the actual number of slow degrees of freedom diminishes. In section \ref{GenerFastSlow} we shall
discuss how, if one concentrates on the slow degrees of freedom, the
detailed balance property has a wider domain of validity. In section
\ref{XorSatExample} we test this with a non-trivial model: constraint optimization of
XorSAT problems with Boolean variables  (we have also tried SAT, with similar results). The purpose here is {\em
not} to argue in favor of the biological relevance of such  models,
but just to test our developments in a highly non-trivial case. The
last part of the paper, section \ref{ExploitCorr}, is dedicated to suggested applications
of this correspondence.

\section{Emergence of thermal dynamics}
\label{EmergThermal}
\vspace{0.5cm}

\textbf{Setting }

\vspace{0.5cm}

We shall consider a population \textcolor{black}{of individuals whose number is} kept constant -- or slowly varying
-- due to the limited resources, such as nutrients or space. For example
a chemostat
regulates the population by
an influx of fresh medium (containing nutrients) and an outflux that
removes medium containing cells. The individuals are assumed to be
independent, except for the indirect interaction that derives from
keeping the total population constant. Each individual is in an internal
state ``i'' (which may describe the genotype and possibly additional
phenotypic traits). It has on average $\lambda_{i}$ offspring per
unit time. In addition, the population is kept constant by introducing
a suitable probability of killing individuals. Specifically, for a
Moran process used below \cite{moran_statistical_1962} the population
size is kept exactly constant, by killing an individual chosen at
random at every event of reproduction. The probability
of mutation per generation a state $i$ to a state $j$ is $\mu_{ij}$,
so that mutation times are random with average $\tau_{ij}=1/\left(\lambda_{i}\mu_{ij}\right)$ .
In the literature, either the probabilities $\mu_{ij}$ or the times
$\tau_{ij}$ are often taken identical for all allowed mutations.
Both options are discussed in the following.

%

\begin{figure}
\centering \includegraphics[width=0.8\textwidth]{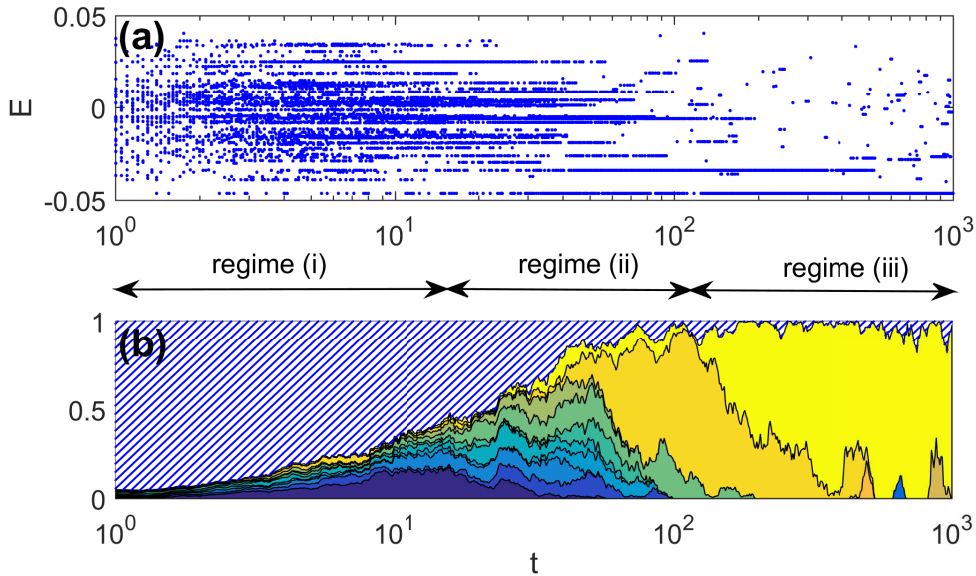}
\protect\protect\protect\protect\caption{Evolution in the REM house-of-card model. (a) The energy of individuals.
For clarity only 1 in 20 individuals is shown. (b) Evolution of the
largest sub-populations, as in Fig. \ref{concu1}. Model parameters
$M=500,\,N=2500,\,\mu=1/50$.}
\label{evo} 
\end{figure}

The evolution is described by a time-dependent distribution of types
$\{n_{1},...n_{2^{N}}\}(t)$, with $\sum_{i}n_{i}(t)=M$. Initial
conditions need to be specified, such as a population containing a
single `wild-type', or a random selection of states for the $M$ individuals.
Qualitatively, one may have several regimes, see Fig. \ref{concu1}.
The system may switch between the regimes as time progresses, as illustrated in Fig.\ref{evo} and discribed below.

{\em i) Continuous population:} essentially all the population
is in states $a$ such that $1\ll n_{a}\ll M$, for all other states
$n_{a}=0$. One may treat the problem in terms of a continuous approximation
$\rho(\lambda)$ corresponding to the fraction of individuals having
$\lambda_{i}$ between contained between $\lambda$ and $\lambda+d\lambda$,
using the Replicator Equation \cite{nowak_evolutionary_2006}. For
example, in the `House of Cards' model, where $\mu_{ij}$ is \textcolor{black}{identical}
for all pairs $i,j$, 
\begin{equation}
\dot{\rho}(\lambda)=[\lambda\left(1-\mu\right)-\langle\lambda\rangle]\rho(\lambda)+\mu\lambda\bar{p}(\lambda)\hspace{1.5cm}with\hspace{1.5cm}\langle\lambda\rangle=\int d\lambda\,\rho(\lambda)\,\lambda\label{evolution-1}
\end{equation}
where $\mu\equiv\sum_{j}\mu_{ij}$ \textcolor{black}{is the total probablity of mutating out of the interval $[\lambda,\lambda+d\lambda]$,}  and $\bar{p}(\lambda)$ is the
density of states.

\vspace{0.1cm}

{\em ii) Concurrent mutations regime }\cite{desai_beneficial_2007}:
Finite population size effects cannot be neglected even if the
population starts at regime (i), because they begin to show up at
times of order $\ln M$. Here a finite fraction of all individuals
are concentrated in a finite number of types, see Figs. \ref{concu1},\ref{evo},
competing for domination (strong clonal interference). (See \cite{desai_beneficial_2007,rouzine_solitary_2003,rouzine_traveling-wave_2008}
and \cite{good_distribution_2012}, especially Refs. 23-29 therein).

\vspace{0.1cm}

{ \em iii) Successional Mutation Regime}: The system settles into
a regime in which the majority of individuals belong to a single type.
Some of these individuals mutate, most often deleteriously, and die,
accounting for a constantly renewed population `cloud' of order $\mu M$
outside the dominant sub-population. Every now and then, an individual
mutates to a state that is more fit, in which case it may spread in
the population until completely taking over (fixation). There are,
in addition, events in which the entire population may get fixated to a mutation that is (slightly) less fit: these extinction events
are exponentially rare in $M$. In this regime, it is easy to compute
the probability for a new mutation to appear and fix in an interval
of time $\delta t$ (large with respect to the fixation time, small
compared to the time between successive fixations) \cite{nowak_evolutionary_2006}
\begin{equation}
P({\mbox{fixed in }}i\to{\mbox{fixed in }}j)\equiv P(i\rightarrow j)
=M\lambda_{i}\mu_{ij}\delta t\;\frac{\frac{\lambda_{i}}{\lambda_{j}}-1}{(\frac{\lambda_{i}}{\lambda_{j}})^{M}-1}
=M\lambda_{i}\mu_{ij}\delta t\; \frac{e^{\left(E_{j}-E_{i}\right)/N_{s}}-1}{e^{\left(E_{j}-E_{i}\right)M/N_{s}}-1}.\label{mc-1}
\end{equation}
where
\begin{equation}
E_{i}\equiv-N_{s}\ln\lambda_{i}\:,\label{eq:E_def}
\end{equation}
is  a quantity that will play a role analogous to that of an energy, and $N_{s}$ is a scale factor that we are free to
choose (units of energy), to make quantities of interest, such are
changes in fitness, of order one.
A population will evolve in this regime whenever mutations are rare
\cite{mustonen_fitness_2010,mustonen_molecular_2008} $\mu M\ll1$,
so that few mutations are offered in any generation.\footnote{{}
If one allows for many mutations to exist, while still having a single
dominant population at almost all times, a somewhat different regime
is obtained \cite{desai_beneficial_2007}. Then Eq. (\ref{mc-1})
no longer holds due to the population `cloud' of deleterious mutations.
If these mutants do not reproduce ($\lambda=0$), Eq. (\ref{mc-1})
may be mended by considering an effective $M_{eff}=M-M_{cloud}$,
but for more general deleterious mutations a simple prescription is
hard to give. However, this correction is small when mutation rates
are low $\mu\ll1$ (but not necessarily very low $\mu M\ll1$). More
precisely, Eq. (\ref{mc-1}) holds when the fraction of deleterious
mutations is small, $\frac{\mu\lambda}{\lambda-\lambda_{del}}\ll1$,
where $\lambda$ is the fitness of the dominant population and $\lambda_{del}$
is a typical fitness of deleterious mutations.}

To make this discussion less abstract, before discussing the more
general situation, let us consider a concrete example where this scenario
is materialized. The mutation rates $\mu_{ij}$ are \textcolor{black}{identical} for all
pairs $i,j$ (the `House of Cards' model), with $\mu=\sum_{j}\mu_{ij}$,
so an individual may jump between any two states. There are states
$i=1,...,2^{N}$ with log-fitnesses distributed according to a Gaussian
distribution (a choice inspired by the Random Energy Model \cite{derrida_random-energy_1980,neher_emergence_2013},
see below)
\begin{eqnarray}
p(E) & = & {\cal N}e^{-E^{2}/2N}\hspace{1cm}\bar{p}(\lambda)=\frac{dE}{d\lambda}\;p(E)
\end{eqnarray}
{We choose the scale factor appearing in Eq. (\ref{eq:E_def})
to be $N_{s}=N$, the logarithm of the number of states, or `genome
length'.}

The evolution is depicted in Fig \ref{evo}, where initially each
individual is chosen randomly. The system traverses through all three
regimes described above, starting from (i), moving to (ii), and finally
reaching regime (iii).

\begin{figure}
\centering \includegraphics[angle=270,width=7cm]{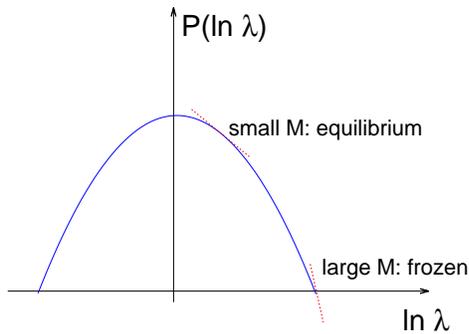} \protect\protect\protect\protect\caption{Equilibrium for the REM/House of Cards model.}
\label{remsketch} 
\end{figure}

\vspace{0.5cm}

\textbf{Detailed Balance in the House-of-Cards model}

\vspace{0.5cm}

For the House-of-Cards model, consider now {\em for the successional
mutation regime} the `meta-dynamics' of the dominant sub-population,
considered as a single entity, neglecting the relatively short times
in which the system is not concentrated into a single type (the {{}
fixation processess}). Since in this example $\mu_{ij}=\mu_{ji}$:
\begin{equation}
\frac{P(i\to j)}{P(j\to i)}=\left(\frac{\lambda_{j}}{\lambda_{i}}\right)^{M-2}=e^{-(M-2)[\ln\lambda_{i}-\ln\lambda_{j}]}=e^{-\frac{M-2}{N}[E_{j}-E_{i}]}
\end{equation}
This corresponds to a process with detailed balance and temperature
$T=\frac{N}{M-2}$ and energies $E_{i}$ \cite{berg_adaptive_2004,berg_stochastic_2003,sella_application_2005}.
If the symmetry is in mutation times $\tau_{ij}=\tau_{ji}$ the temperature
becomes $T=\frac{N}{M-1}$. In what follows we focus on large $M$,
and dropping $O\left(1/M\right)$ corrections we write $T=\frac{N}{M}$.
At very long times, the system will reach a distribution 
\begin{equation}
P({\mbox{dominant type at }}i)=\frac{e^{-E_{i}/T}}{\sum_{j}e^{-E_{j}/T}}\label{equil}
\end{equation}

Still in the present model, finding the stationary distribution has
been reduced to the solution of the \emph{equilibrium} Random Energy
Model \cite{derrida_random-energy_1980}, (see \cite{neher_emergence_2013}
and \cite{cammarota_spontaneous_2014}). In particular, we conclude
that, depending on the value of $M$, at very long times the system
will equilibrate to either a liquid phase (for $M/N<\ln2$) or a `frozen'
phase (for $M/N>\ln2$), see Fig. \ref{remsketch}: in the former random
extinction events stop the system from converging to the optimum level
of fitness, while in the latter this level is at long times reached.
{\em A feature we find here, and is a general fact, is that even
if the dynamics satisfy detailed balance, and are hence able in principle
to equilibrate, this takes place at unrealistically long times.}

The qualitative features of the population at different times has
long been known, the similarity of the role played by fluctuations
due to finite population size with thermal fluctuations has also been
noted long ago \cite{peliti_introduction_1997,crow_introduction_1970}.
Here the analogy becomes an identity, and features such as Muller's
ratchet -- the accumulation of deleterious mutations in an irreversible
manner -- become just the question of an ordinary order-disorder phase
transition. Similarly, the effect of population bottlenecks becomes
the same as a spike in temperature.

\vspace{0.5cm}

\textbf{Detailed Balance for non-symmetric mutation rates}

\vspace{0.5cm}

Before concluding this section, let us note the result, valid for
more general $\mu_{ij}$, but still satisfying (\ref{mc-1}) . In that case, the condition for detailed
balance is \cite{berg_adaptive_2004,berg_stochastic_2003} 
\begin{equation}
1=\frac{P(i\to j)P(j\to k)P(k\to i)}{P(k\to j)P(j\to i)P(i\to k)}=\frac{\mu_{ij}\mu_{jk}\mu_{ki}}{\mu_{kj}\mu_{ji}\mu_{ik}}\label{detba}
\end{equation}
 meaning that the system satisfies detailed balance
if the mutation probabilities $\mu_{ij}$ also do. The detailed balance
condition (\ref{detba}) for them may be written: $\mu_{ij}=\hat{\mu}_{ij}e^{-\beta_{\tau}\hat{E}_{i}}$
for some $\hat{E}_{i}$ and $\hat{\mu}_{ij}=\hat{\mu}_{ji}$, and
temperatures and energies become 
\begin{equation}
T=\beta^{-1}=\frac{N_{s}}{M}\hspace{1cm}E_{i}=-N_{s}\ln\lambda_{i}+\frac{\beta_{\tau}}{\beta}\hat{E}_{i}\label{eandt1}
\end{equation}
Throughout this paper we shall consider symmetric mutations $\mu_{ij}=\mu_{ji}$,
so that $\beta_{\tau}=0$ and we recover $E_{i}=-N\ln\lambda_{i}$.

If a system evolves under a dynamics satisfying the detailed balance
condition, it will after long times equilibrate to a distribution:

\begin{equation}
P_{eq}(i)={\cal N}e^{-\beta E_{i}}={\cal N}\;\lambda_{i}^{M}\;e^{-\beta\hat{E}_{i}}
\end{equation}
the last factor being absent when $\mu_{ij}=\mu_{ji}$.

\vspace{0.5cm}

\textbf{Limitations }

\vspace{0.5cm}

The possibility of mapping evolutionary models to systems evolving
in contact with a thermal bath seems very appealing as all the artillery
of statistical mechanics: fundamental inequalities, analytic and numerical
methods, fluctuation relations,  immediately becomes available.
It is for this reason surprising that the approach has received less
attention than it deserves, even amongst physicists. This is probably
due to a number of reasons, most importantly:

\vspace{0.2cm}

$\bullet$ The entire construction relies on a timescale-separation
allowing for complete clustering (fixation) into a single-state population.
In other words, the system has to be in the successional mutation
regime.

$\bullet$ In the successional mutation regime, even if (meta)dynamics
allows for equilibration, this will not realistically be reached before
astronomical times. An equation like (\ref{equil}), and all the ones
relying on it, will then not be applicable. In other words, {\em
the dynamics may be willing to equilibrate, but the times are short.}
The mapping will be at best with an out of equilibrium system in contact
with a thermal bath.

As we shall see in what follows, the second objection brings in a
solution of sorts to the first objection: because we have to necessarily
deal with systems that are out of equilibrium at long times, a timescale-separation
between processes that are fast and slow appears naturally (and inevitably).
This separation may lead to detailed-balance, when generalized appropriately.

\section{A natural source of timescale-separation: the glassy element }
\label{TimescaleSep}

In this section we briefly evoke  the following fact: {\em any long-time out
of equilibrium system (as evolutionary systems will inevitably be)
necessarily has fast and slow processes, the timescale separation
growing with time.}

Some physical systems do not equilibrate in experimental times, even
if placed in contact with an equilibrium thermal bath. In practice,
this means that equilibrium concepts such as temperature ({\em of
the sample}) cannot be applied: two thermometers may measure different
values depending on the way they are coupled to the system. On the
other hand, the sample has an {\em age}, the time since it was
prepared, that may affect its physical properties.

Three examples help clarify the different possibilities leading to
time-scale separation. Consider first the case of diamond: it is actually
a metastable state at room temperatures and pressures, there is always
a probability of a sample decaying to graphite - the stable form of
carbon. Despite its metastable, and hence nonequilibrium nature, while
diamond lasts it behaves very much as an equilibrated system, only
the decay itself would tell us otherwise. Next, consider the case
of crystal ripening: the sample is constituted by an assembly of micro
crystallites. Slowly but continuously, these grow one at the expense
of the others, a process that only stops when the entire sample is
a monocrystal. Nonequilibrium processes (the motion of crystal walls)
are continuously happening, although ever more slowly. These processes
coexist with the high-frequency vibrations within crystallites, which
are essentially those of an equilibrium sample. Finally, consider
the examples of glasses. These have no obvious ordered structure,
but are constantly evolving to more and more equilibrated configurations.
Once again, the high frequencies are just those of an equilibrated
solid, while the slow processes are non-equilibrium, although in this
case we have as yet no obvious way to picture them as we had with
the example of crystallite growth.

\begin{figure}
\centering \includegraphics[angle=270,width=8cm]{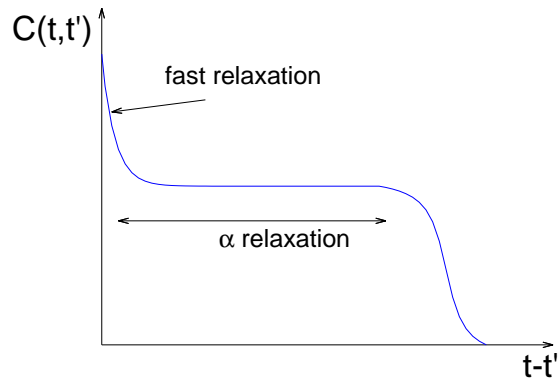} \protect\protect\protect\protect\caption{Timescale-separation}
\label{c} 
\end{figure}

In any one of these systems, we may define an {\em autocorrelation
function} $C(t,t')$, that measures how much the system evolves.
It is defined to be $C(t,t')=1$ if and only if the configurations
are the same at $t$ and $t'$, and falls to zero if they are completely
uncorrelated. (We shall define a concrete $C(t,t')$ for a model below).
The fast and slow processes described above manifest themselves in
a form of the autocorrelation as shown in Figure (\ref{c}). There
is a fast fall of the correlation to a plateau, corresponding to the
vibrations that are insensitive to the out of equilibrium nature.
At longer times, there is the slow relaxation ($\alpha$-relaxation,
in the glassy jargon), that corresponds to the slow degrees of freedom.
In long-time out of equilibrium systems the $\alpha$-relaxation time
$t_{\alpha}$ becomes longer and longer as time passes, while the
fast relaxations are rather insensitive to the overall time elapsed
since preparation. 

This fact (fast relaxations plus slower and slower $\alpha$ relaxations),
is a completely generic feature, that one may consider true by definition:
if there were only fixed timescales the system would be unchanged
in time, i.e. in equilibrium. If we wish to picture this as a motion
in phase space, two explanations are often given:

{\em i)} States: the system jumps discontinuously between valleys.
While it is in a valley, its fluctuations are those of equilibrium,
they are indifferent to the fact that the present valley is not the
optimal one. It takes a time $t_{\alpha}$ to jump from a valley to
another. Because the system optimizes, it becomes harder and harder
to find new, deeper valleys, and $t_{\alpha}$ grows.

{\em ii)} Canyon: a rather abstract, but in fact more relevant
mechanism in systems with many degrees of freedom is understood by
considering the system evolving in a `canyon', with many transverse
directions that are essentially in equilibrium, while there is a slow
drift along the canyon. This drift becomes slower and slower as time
passes, partly because the dimensionality of the `canyon' diminishes continuously -- the number of `almost flat' directions becomes smaller and smaller. This is exactly what happens in the ripening process, the
slow directions being the coordinates that parametrize the (slowly disappearing) crystallite
walls. This example is important, because it stresses the fact that
fast and slow processes may happen simultaneously \cite{kurchan_phase_1996}.

We have already met timescale-separation in the previous example of
the House of Cards model. Even if the mutation timescale $\tau$ is
not very large, a time comes when the evolution slows down because
it is very unlikely for a mutation to become favorable, and the population
becomes essentially monoclonal (except for a `cloud' of deleteriously
mutated individuals, that itself depends on $\tau$). This slowing
down due to a progressive difficulty in optimizing further is itself
closely analogous to the situation in glassy systems. {In
the system with isolated states (Golf-course) one has that jumps become
rarer and rarer, and hence dynamics slow on average, but a successful
jump itself, when looked retrospectively, takes a short time. In this
sense, the } "canyon'' picture is quite different:
the dynamics becomes slower, but there are, most of the times, no
fast events. In fact, complex systems such as glasses exhibit the
latter, rather than the former behavior. This will turn out to be
relevant for our discussion here.
Bearing these facts in mind, we now turn back to population dynamics.

\section{Generalization to problems with fast and slow timescales: three examples }
\label{GenerFastSlow}

Aging systems automatically generate fast and slow timescales. For
an aging evolutionary system, fast timescales  will invalidate the
approach that leads to detailed balance: there can be no clustering
if there are fast successful mutations\textcolor{black}{, that compete for fixation simultaneously}. 
On the other hand, we expect that the
slow timescales might yield some form of clustering, and hence of
detailed balance, provided we coarse-grain over the short times. This
section contains exercises to show that this is possible.

\vspace{0.5cm}

\textbf{Successional regime in a picture with barrier crossing between `super-states'}

\vspace{0.5cm}

%

We may first attempt to model a 
 situation where there are   `super-states', each containing many states.  Jumps
between super-states become more and more infrequent, but there is no clustering
in single states.   The separation between timescales may  come about  in at least two ways:

$\bullet$ Because mutation probabilities  probabilities are large between states belonging to the same  super-state, and small 
otherwise.

$\bullet$ Because mutations between states within a superstate bring about changes of fitness that are much smaller
than those allowed between different superstates.

Here we take as an example the former possibility.
We consider the House of Cards just as before, but
this time with  `super-states' $i$ with $i=1,...,2^{N}$, and, within each, 
 states $(i,a)$ with $a=1,...,K_{i}$. The distribution of fitnesses
is $\lambda_{ia}$. The timescale for jumps between states
of different super-states $(i,a)\to(j,b)$ (with $i\neq j$) are large
numbers $\tau_{ij}$, independent of $a$ and $b$. The timescale
of jumps between configurations within a state $(i,a)\to(i,b)$ are
much shorter $\tau_{o}\ll\tau$ .
We shall assume that conditions are such that the population is most
of the time clustered in a single super-state, but within a super-state
it evolves fast (i.e. jumping fast between $(i,a)$ and $(i,b)$)
and we may assume that the population is distributed according to
a distribution $p_{ia}$. (This distribution need not, as we shall
see, be the one corresponding to $M=\infty$, but it is not a thermal
distribution either. In practice, it has to be calculated from first
principles for given $M$). The total rate of reproduction within
a super-state $i$ is then: 
\begin{equation}
\bar{\lambda}_{i}\equiv\sum_{a}p_{ia}\lambda_{ia}
\end{equation}
Assuming that rearrangements inside a super-state are rapid with
respect to the time for a fixation from one super-state to another,
we may take the average rates $\bar{\lambda}_{i}$.  

It is easy to see that, under these conditions, detailed balance holds then for jumps between super-states. The role
of fitness  is played by an averaged value within each super-state.
 On the other hand, within a super-state jumps are
fast and there is no clustering: the whole notion of thermal dynamics
does not apply for this regime.

\vspace{0.5cm}

\textbf{Diffusion in smooth fitness landscape }

\vspace{0.5cm}

 In the previous section, we described two explanations
for the slow-down of dynamics. In explanation (i) above, slow dynamics
result from motion in valleys, separated by barriers that are rarely
crossed. Such an evolution corresponds to a Successional Mutation Regime,
with additional mutations inside the valleys. The existence of detailed
balance in this case, as well as the limitations on its validity,
were illustrated with the example above. We now turn to discuss evolutionary
dynamics as follows from explanation (ii), in which  fast and slow evolutions 
coexist, and are not necessarily separated as in the case of discrete
jumps between states. The problem we consider was treated in the
literature twenty years ago: in a remarkable paper \cite{kessler_evolution_1997}
all the details have been worked out, but no mention was made of how
this implied a detailed balance relation in agreement with the one
discussed above.
\begin{figure}
\centering \includegraphics[scale=1.5]{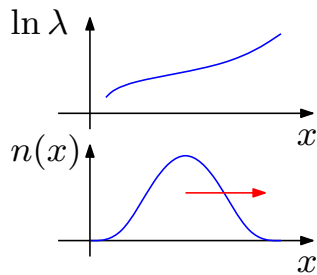} \protect\protect\protect\protect\caption{Traveling population pulse. $n\left(x\right)$ is the population distribution.
Red arrow shows direction of average motion.}
\label{populationpulse} 
\end{figure}

Consider a population of $M$ individuals performing simple diffusion
along $x$ with diffusivity $D$, and reproducing (or dying) with
a rate $\lambda(x)=e^{-N_{s}^{-1}E(x)}$ \cite{kessler_evolution_1997}.
The offspring diffuses with a different realization of noise \footnote{{}The
continuous model is in the spirit of \cite{zhang_diffusion_1990,meyer_clustering_1996,kessler_evolution_1997}.
In fact, Kessler \emph{et. al.} considered a situation in which $x$
is discretized. Since they work in the regime in which the `pulse'
is wide with respect to the discretization, the continuous limit is
easily obtained from their construction.}. It is clear that under
such conditions no two individuals are in exactly the same position,
except at the precise moment of birth. In fact, the population takes
the form of a pulse of width $\Delta x=\sqrt{\frac{D}{\lambda}M}$,
{which represents a balance between spreading due
to diffusion, and narrowing due to replication events.} The pulse's
center of mass diffuses with the diffusivity $D_{pulse}$, and drifts
with speed $=\kappa_{pulse}N_{s}\nabla\ln\lambda=-\kappa_{pulse}\nabla E(x)$
in the limit in which the variations of $E(x)$ are slow. For time-differences
much longer than the relaxation time of the pulse, the system can be shown to  obey
a Langevin equation 
\begin{equation}
\dot{x}=-\kappa\nabla E(x)+\eta(t)
\end{equation}
where $\eta(t)$ can be taken to be a white noise $\langle\eta(t)\eta(t')\rangle=D_{pulse}\delta(t-t')$.
For very small fitness gradients (see condition below), both $D_{pulse}$
and $\kappa_{pulse}$ have been calculated years ago. It was shown
that \cite{zhang_diffusion_1990,meyer_clustering_1996,kessler_evolution_1997}
$D_{pulse}=D$, and that \cite{kessler_evolution_1997} $\kappa_{pulse}=\left(M/N_{s}\right)D_{pulse}$,
a fluctuation-dissipation relation with a corresponding inverse temperature
$T^{-1}=M/N_{s}$. Remarkably, the dynamics satisfy detailed-balance
exactly as in the very different regime of successional mutations,
{{\em}and with the same temperature\textcolor{green}{}}\emph{,}
$T=M/N_{s}$. Here many concurrent mutations are allowed, thus relaxing
the condition of successional mutations.

{This detailed balance property holds if changes
in log-fitness $\Delta\left(\ln\lambda\right)$ of an individual in
a generation are small compared to $1/M$, or $\sqrt{D/\lambda}\nabla E(x)\ll N_{s}/M$.
Noting that $\frac{d}{dt}\langle E\rangle=\frac{M}{N_{s}}D(\nabla E)^{2}$
we can rewrite this condition as $\frac{1}{\lambda}\frac{d}{dt}\langle E\rangle\ll N_{s}/M$,
which is precisely a timescale separation -- rate of average fitness
improvement {\em versus} reproduction rate.}

Unfortunately, the condition on the smallness of the gradient is quite
restrictive. Indeed, the selective advantage (increment in log-fitness)
of a single mutation must be smaller than $1/M$, which is unrealistic
in many problems of cellular evolution. Outside this regime, for stronger
fitness gradients, another regime is entered, studied in detail by
Desai and Fisher, and by Rouzine \emph{et al.} \cite{desai_beneficial_2007,rouzine_solitary_2003,rouzine_traveling-wave_2008}.
When this regime holds, detailed balance is lost. As we now discuss,
in problems with slow evolution in a high-dimensional rugged landscape,
detailed-balance survives to regimes that contain both concurrent
mutations, and larger fitness increments in a single mutation. 

\vspace{0.5cm}

\textbf{{Smooth fitness landscape within a canyon}}

{\vspace{0.5cm}
}

{The model discussed above is a starting point for
a `canyon' model, where the slow direction in a canyon picture varies
slowly and smoothly, while the fitness rapidly decreases in transverse
directions. In this multi-dimensional model there are now two conditions
for detailed balance to hold. First, in the slow direction we need
$\frac{1}{\lambda}\frac{d}{dt}\langle E\rangle\ll N_{s}/M$ as in
the one dimensional model. Secondly, the `cloud' of deleterious mutations
caused by mutation in fast directions must be small. (As for the successional
regime, the condition is $\frac{\mu\lambda}{\lambda-\lambda_{del}}\ll1$,
where $\lambda,\lambda_{del}$ are the fitnesses in the pulse and
the deleterious mutations, respectively.) Note that here the discreteness
of the mutations in the fast directions is important, and cannot be
replaced by a smooth landscape.}

\vspace{0.5cm}

{\bf Detailed balance after temporal coarse-graining.}

\vspace{0.5cm}

In the `super-state' model, in smooth landscape picture, and its extension
to a multi-dimensional `canyon' model, detailed balance does not hold
between individual states, as these change rapidly and in a non-thermal
way. Instead, it holds for states when properly averaged over the
population and a suitable time window. Let us see how this comes about.
We consider  a coarse-grained situation with ensembles of states  $i$, with an average
fitness given  by $\bar{\lambda}_{i}$:
\begin{equation}
\bar{\lambda}_{i}\equiv\int_t^{t+\Delta}  dt\sum_{a}\lambda_{a}\left(t\right)
\end{equation}
where the sum runs over the population and time integrated over a time
window $\Delta$ that is long compared to rapid fluctuations, but short compared
to residence time in a super-state. In the smooth landscape model
(and canyon model) the Einstein relation $\kappa_{pulse}=\left(M/N_{s}\right)D_{pulse}$
implies that between super-states 
\begin{equation}
P(i\to j)=\tilde{\mu}_{ij}\;\frac{e^{\left(F_{j}-F_{i}\right)/N_{s}}-1}{e^{\left(M/N_{s}\right)(F_{j}-F_{i})}-1}\label{mc1-1}
\end{equation}
where $F_{i}\equiv-N_{s}\ln\bar{\lambda}_{i}$. Note that $F_i$ contains also the logarithm of the multiplicity, i.e. an entropic factor.  Detailed balance then
holds with
\begin{equation}
\frac{P(i\to j)}{P(j\to i)}=e^{-\left(M/N_{s}\right)\left(F_{j}-F_{i}\right)}
\end{equation}

\begin{figure}
\centering \includegraphics[width=0.4\textwidth]{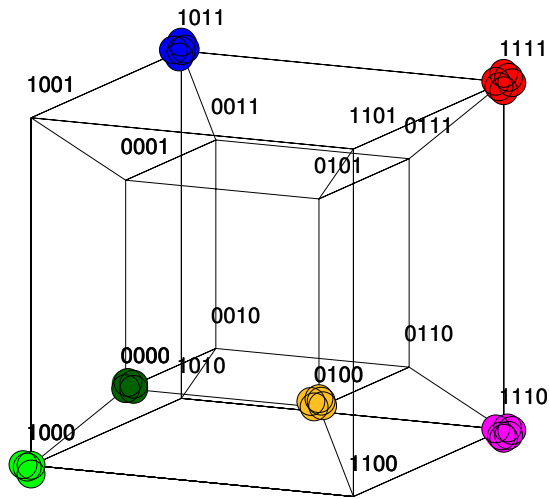} \protect\protect\protect\protect\caption{Genetic algorithm for SAT or XORSAT on the 4-dimensional hypercube. \textcolor{black}{Each vertex has a given fitness value. Individuals reproduce with the rate determined by the vertex they are in, and mutate by diffusing to connected vertices.}}
\label{cube} 
\end{figure}

{Detailed balance holds then for jumps between super-states.
The role of energy is played by an averaged value within each super-state,
together with the number of states within a super-state, leading to
the free energy. On the other hand, within a super-state jumps are
fast and there is no clustering: the whole notion of thermal dynamics
does not apply. Given very long times, the system will eventually
equilibrate. By this we mean that 
\begin{equation}
P(clustering\;within\;super-state\;i)={\cal {N}}e^{-MF_{i}}\label{uqui-1}
\end{equation}
Let us emphasize  that the distribution within this cluster is {\em
not} given by a Boltzmann factor, and requires the full solution
of the population dynamics within the super-state.

\vspace{0.2cm}

{\bf A Fluctuation Theorem}

\vspace{0.2cm}

Consider a quantity $A_{ia}$ defined for each state. The average
over a super-state $\bar{A}_{i}$ may be expressed as an average over
a time $t_{in}$ sufficient to explore inside a state but short to
move between super-states. 
\begin{equation}
\bar{A}_{i}\sim\frac{1}{t}_{in}\int_{t}^{t+t_{in}}\;dt\;A_{ia}(t)\label{window-1}
\end{equation}
We also define $F_{i}^{h}=F_{i}-h\bar{A}_{i}$.

{Assuming that the initial probability distribution
{\em of super-states} is equilibrium with $h=0$ (Equation (\ref{uqui-1}))
$P(i)\propto e^{-\beta F_{i}}$, we switch the field and wait for
a time that is long with respect to the redistribution time inside
a super-state, but otherwise arbitrary. The probability distribution
of a change in the value of $\bar{A}$ is then $P(\Delta)$ which
satisfies a Fluctuation Theorem 
\begin{eqnarray}
P(\Delta) & = & \sum_{ij}e^{-\beta F_{i}}P(i\to j)\delta[\Delta-(\bar{A}_{i}-\bar{A}_{j})]=\sum_{ij}e^{-\beta F_{i}^{h}}P(i\to j)e^{-\beta h\bar{A}_{i}}\delta[\Delta-(\bar{A}_{i}-\bar{A}_{j})]\nonumber \\
 & = & \sum_{ij}e^{-\beta F_{j}^{h}}P(j\to i)e^{-\beta h\bar{A}_{i}}\delta[\Delta-(\bar{A}_{i}-\bar{A}_{j})]=\sum_{ij}e^{-\beta F_{j}}P(j\to i)e^{-\beta h(\bar{A}_{i}-\bar{A}_{j})}\delta[\Delta-(\bar{A}_{i}-\bar{A}_{j})]\nonumber \\
 & = & e^{\beta h\Delta}P(-\Delta)\label{ft-1}
\end{eqnarray}
Note that this is valid even before the system had time to re-equilibrate
at the new value of $h$, but it does require that there was equilibrium
at $h=0$ at the outset. This is why the usefulness of this equation
is rather limited in real life, although we will use it as a proof
of principle in the next section.}

\section{A model example}
\label{XorSatExample}

In this section we investigate how the above considerations apply
to a model for which timescale-separation is not a feature put in
`by hand', but one that arises naturally. Our purpose is not to propose
this model as a useful metaphor (there are many references on this,
see for example \cite{pedersen_long_1981,sibani_evolution_1999,seetharaman_evolutionary_2010,saakian_eigen_2004,saakian_evolutionary_2009,saakian_solvable_2004}),
but rather work out the details in a nontrivial case. A complete analytic
solution for the population dynamics in these models is perhaps possible,
but seems like a daunting task.

We represent the internal state of the cell using Boolean variables.
The fitness is a function on these  variables, and the dynamics
are a Moran process with selection and mutations. The fitness functions
used are standard spin-glass benchmarks, whose landscape properties
have been extensively studied \cite{kauffman_metabolic_1969}. 
The individual's state is defined by  ${\bf s}=\{s_{1},...,s_{N}\}$  variables
taking values $0,1$. Fitness is constructed as follows: there are
$\alpha N$  clauses $a$ with $K=3$ variables, of the form
$(s_{i_{1}^{a}}\vee\overline{s_{i_{2}^{a}}}\vee s_{i_{3}^{a}})$ where
both the $(i_{1}^{a},i_{2}^{a},i_{3}^{a})$ chosen for each clause
-- and the fact that the variable is negated or not -- are decided
at random once and for all. For example, Random K-SAT and Random Xor-SAT
take the form:

\begin{center}
OUTPUT = $(s_{18}\vee\overline{s_{3}}\vee s_{43})\wedge(s_{1}\vee s_{45}\vee\overline{s_{31}})...\wedge(\overline{s_{51}}\vee s_{7}\vee\overline{s_{8}})$
\emph{~~(SAT)}
\par\end{center}

\begin{center}
~~~~~~OUTPUT = $(s_{18}\veebar\overline{s_{3}}\veebar s_{43})\wedge(s_{1}\veebar s_{45}\veebar\overline{s_{31}})...\wedge(\overline{s_{51}}\veebar s_{7}\veebar\overline{s_{8}})$
\emph{~~~(XorSAT)}
\par\end{center}

If we assume that each clause has a multiplicative effect on the reproduction
rate $\lambda$ , this takes us to an additive form for $E$

\begin{equation}
-\ln\lambda=\frac{1}{N}\sum_{a=1}^{\alpha N}[\mbox{error in clause }a]=\frac{1}{N}E\label{ftn}
\end{equation}

The factor $\frac{1}{N}$ sets the energy scale, that is we choose
the scale $N_{s}=N$ (as in the House-of-cards example). Mutation
time-scale $\tau_0=\left(\mu\lambda\right)^{-1}$ is taken equal for
all mutations in a given simulation, and accordingly we make use of
this variable throughout this section.

This problem is a standard benchmark of optimization theory, and has
been extensively studied. We shall work in a regime with $\alpha=6$:
for such a number of clauses the system virtually never has a solution
where all clauses are satisfied, i.e. $E>0$. The landscape is rugged
and the minima are separated and extremely hard to find.

We consider the dynamics of $M$ individuals, each identified by a vector ${\bf s}^{individual}$,
performing diffusion by flipping randomly one of their components,
reproducing or dying with fitness given by the SAT or XorSAT fitnesses.
In other words, our `cells' perform diffusion on the vertices of an
$N$ dimensional hypercube (Fig \ref{cube}), where they reproduce
or die.

To obtain a meaningful phase-diagram (Fig \ref{phasediagram}),
the scaling of $N_{s}$ and $\tau_0$ with growing $M$ must be consistently
defined. We keep $M/N$ constant. To maintain the internal energy
$E/N=-\ln\lambda$ unchanged, numerics show that mutation times must
scale as $\tau_0=\tau M$, where $\tau$ is constant, for a given value of $M/N$. This entails
that the width of the fitness distribution in the population at a
given time is $\sigma_{\lambda}^{2}\sim\frac{M}{N_{s}^{2}\tau_0}\sim\frac{1}{M^{2}}$
(following arguments as in \cite{kessler_evolution_1997,ridgway_evolution_1998} }). The point-mutation time (i.e., time-scale for a given spin) is $\tau_{point}=N\tau_0$. \textcolor{black}{This corresponds also to the time-scale for an individual to shuffle its entire genome.}

\begin{figure}
\centering \includegraphics[angle=270,width=0.6\columnwidth]{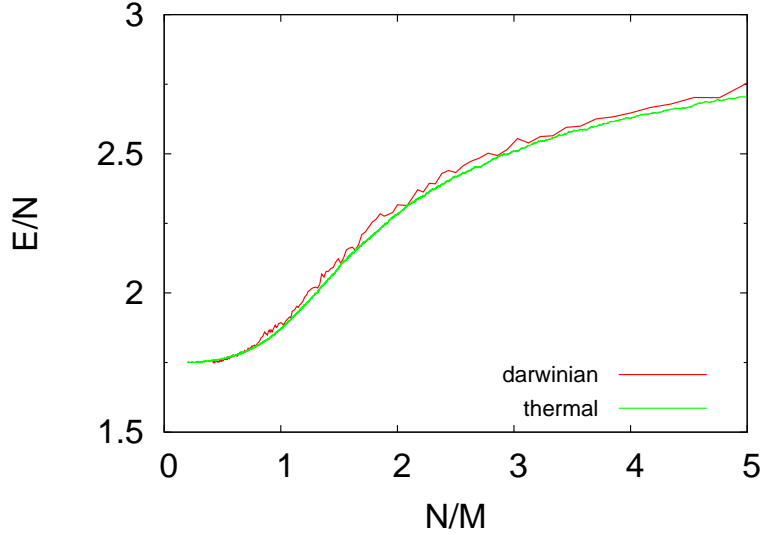}
\protect\protect\protect\protect\caption{Thermal versus Darwinian annealing for XorSAT, for $\tau=16$. Similar
results are obtained for K-SAT. Darwinian annealing is performed by
controlling the population so that it increases slowly, at the same
rate as in the corresponding thermal annealing.}
\label{annealing} 
\end{figure}

Under conditions of the successional regime ($M/\left(\lambda\tau\right)\ll1$),
detailed balance would hold with temperature $T=\frac{N}{M}$. Indeed,
in Fig. \ref{annealing} we show the results of a simulated annealing
performed with an ordinary Monte Carlo program on a single sample,
superposed with a `populational annealing' performed by slowly increasing
the population of a set of individuals performing diffusion and reproducing
according to the fitness in Eq. (\ref{ftn}). Although this is not
the main purpose of this paper, we note in passing that this `thermal'
analysis allows one to make an evaluation of such `genetic algorithms'
-- in this case we understand that the Darwinian Annealing will have
the same strengths and weaknesses as has Simulated Annealing. Furthermore,
we see that allowing for large populations from the outset may be
as catastrophic as is a sudden quench in an annealing procedure. 

\textcolor{black}{In the limit $\tau\rightarrow \infty$, the population is fully clustered, and it behaves like a single realization of a SAT system, at temperature $T=\frac{N}{M}$.}
From what we know of the usual thermal XorSAT or SAT problem \cite{krzakala_gibbs_2007},
there is a (dynamic) glass transition below a certain temperature
$T_{d}$. Below $T_{d}$ the phase-space breaks into components, and
optimization becomes very hard.  (This transition happens before the thermodynamic one, which itself 
is closely analogous to the freezing one of the REM).  We may locate this
transition by plotting the autocorrelation functions $C(t,t')=\frac{4}{N}\sum_{i}(s_{i}\left(t\right)-\frac{1}{2})(s_{i}\left(t'\right)-\frac{1}{2})$
at decreasing temperatures. As $T_{d}$ is approached from above,
the correlation decays in a two-time process, a fast relaxation to
a plateau followed by a much slower $\alpha$-relaxation taking a
time $t_{\alpha}$. As $T_{d}$ is reached, $t_{\alpha}$ diverges.
Below $T_{d}$, the system {\em ages}: the time $t_{\alpha}$ now
keeps increasing with time, $C(t,t')$ decays in a time $(t-t')_{decay}\sim t_{\alpha}(t')$
with $t_{\alpha}(t')$ an increasing function of $t'$.

\begin{figure}
\centering \includegraphics[angle=270,width=0.48\columnwidth]{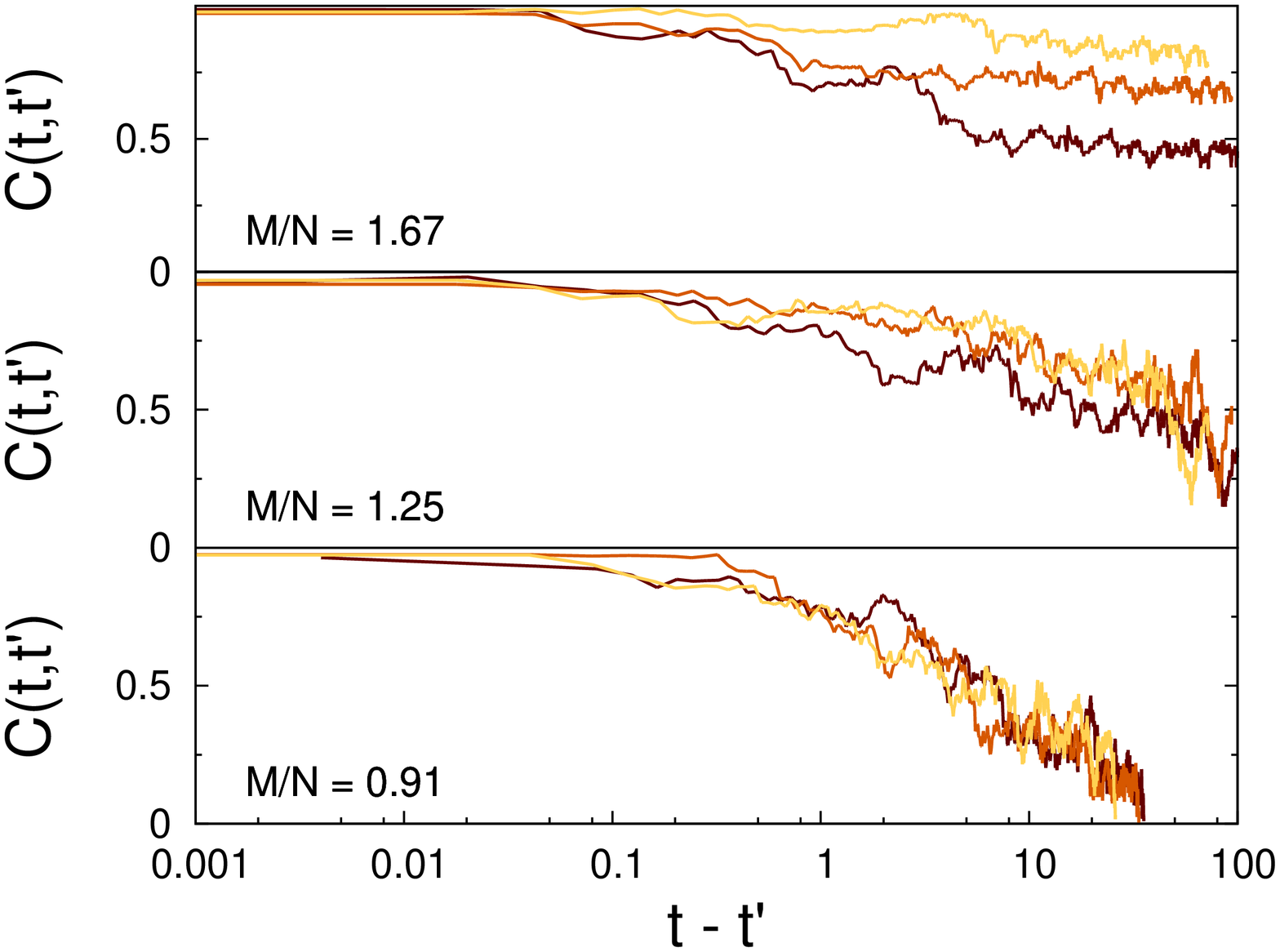}\includegraphics[angle=270,width=0.48\columnwidth]{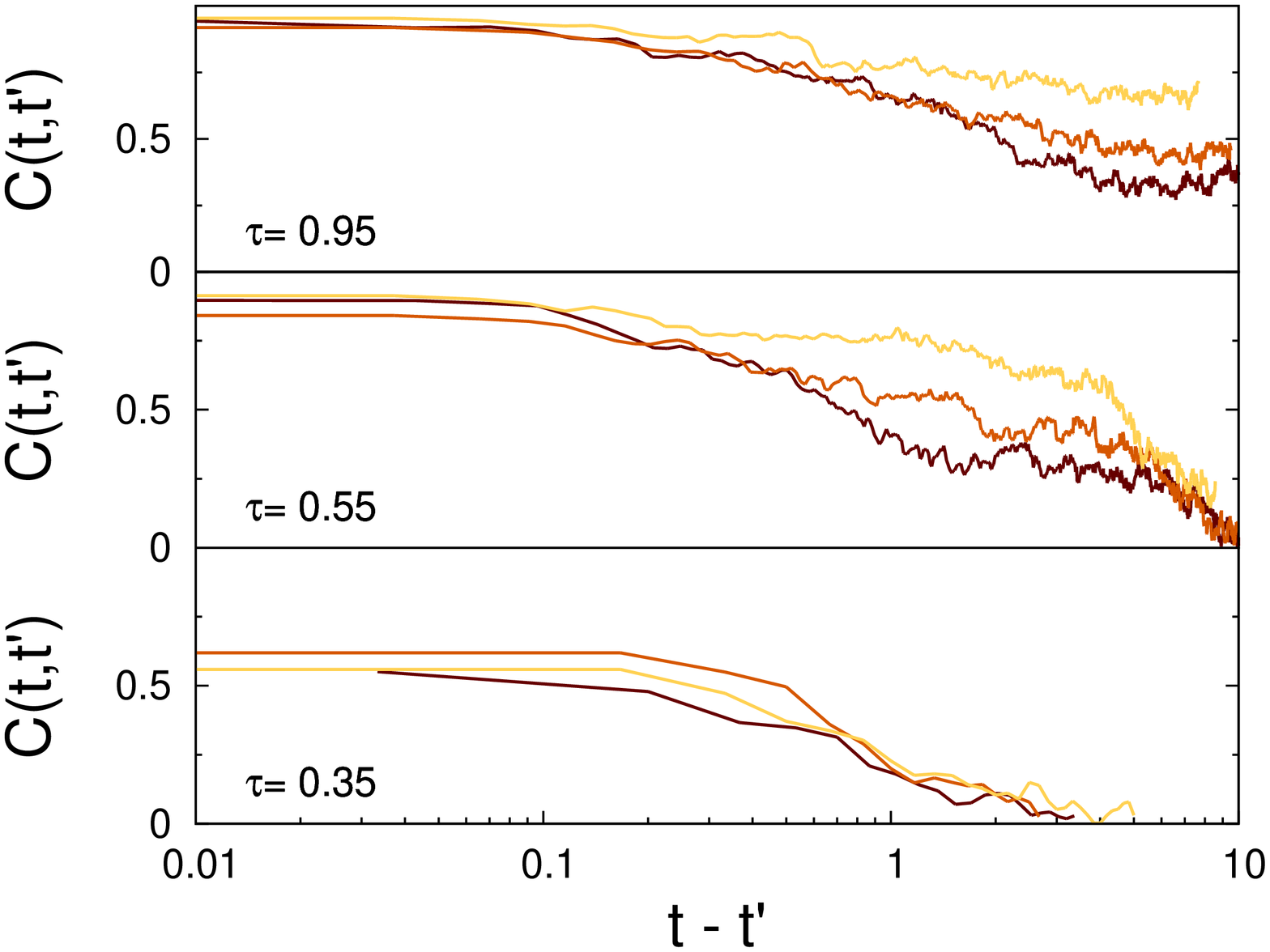}
\protect\protect\protect\protect\caption{Values of the $C(t,t')$ for a population crossing the glass transition
from liquid (bottom) to glass (top). Values are $\tau=5$, $M/N={0.91,1.25,1.67}$
(left), and $M/N=4$, $\tau={0.35,0.55,0.95}$. The transition line
is thus crossed by increasing $M$ or $\tau$, see the two arrows
in fig. \ref{phasediagram}, the points corresponding to the figures
are marked by crosses. The color code indicate the value of $\ln(t')$,
growing linearly from darker to lighter curves. In the top figures
$t_{\alpha}$ grows as the system ages. Here and in all the following
figures, the time is measured in units of (cell-)mutation times $\tau$
(i.e. a cell performs on average $O(N)$ flips in $\Delta t=O\left(1\right)$).
{The correlations are reasonably smooth, for a big
system, }{\emph{even for a single run}}{.
This is evidence for a canyon, rather than ``fast, rare jump'' picture.}}
\label{autocorr} 
\end{figure}

\begin{figure}
\centering \includegraphics[angle=270,width=0.6\columnwidth]{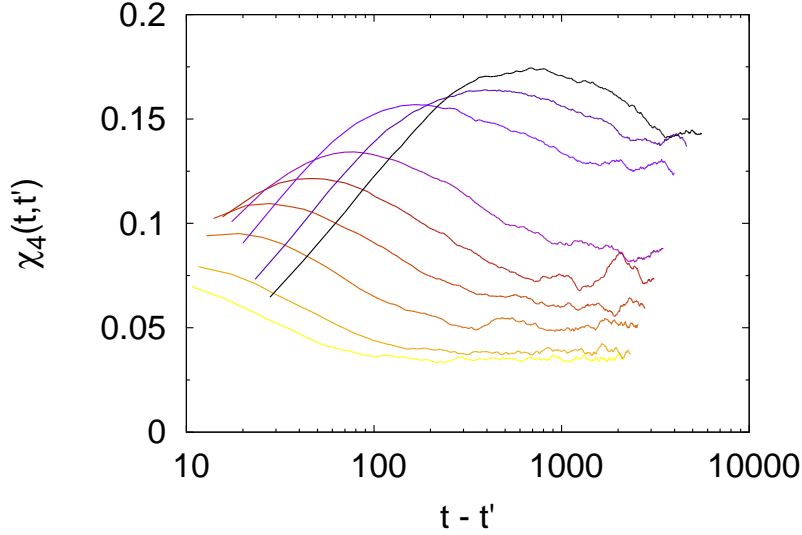}
\protect\protect\protect\protect\caption{$\chi_{4}$ versus time on approaching the transition. Data shown
for $M/N=1.4$, and a mutation rate $\tau^{-1}$ increasing linearly
from 0.25 to 0.65 (bottom to top, dark to light curves). At lower
mutations rates (not plotted) the glassy phase is reached, and the
curves do not present any maximum and keep increasing with the waiting
time $t$, as $t_{\alpha}$ is no longer defined.}
\label{chi4} 
\end{figure}

\begin{figure}
\centering \includegraphics[angle=-90,width=0.5\columnwidth]{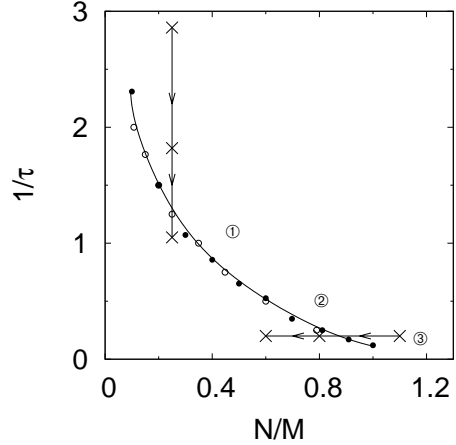}
\protect\protect\protect\protect\caption{Phase transition diagram in the $N/M$ vs $1/\tau$ plane. The transition
line is given by the black circles, which correspond to the points
where the value of the maximum in the $\chi_{4}$ function diverges,
approaching the line by changing $M$ (filled circles) or $\tau$
(empty circles). The two arrows correspond to two possible crossing
of the transition line, changing $\tau$ or $M$, the points indicated
by a cross being those represented in Fig \ref{autocorr}. Points
labeled by numbers correspond to the situations of Fig \ref{concu1},
and indicate the parameters used in the verification of the \emph{averaged}
detailed balance, Fig \ref{FT}.}
\label{phasediagram} 
\end{figure}

What we have described is the `Random First Order Transition' \cite{kirkpatrick_scaling_1989}.
Nothing new here, as the system is equivalent to a thermal system,
known to exhibit such a transition. Let us now consider smaller $\tau$,
so that we no longer can assure that the $M$ individuals are fully
clustered in a configuration at most times. We approach the transition
by increasing $M$ at fixed $\tau$, and by decreasing $\tau$ at
fixed $M$. The correlation curves obtained are shown in Fig (\ref{autocorr}),
the nature of the transition remains the same, but the transition
value of critical $M$ shifts with $\tau$. 
A confirmation of this is obtained by plotting the autocorrelation
`noise' 
\begin{equation}
\chi_{4}=N\left\langle \left[\frac{4}{N}\sum_{i}(s_{i}\left(t\right)-\frac{1}{2})(s_{i}\left(t'\right)-\frac{1}{2})\right]^{2}\right\rangle -NC(t,t')^{2}
\end{equation}
a quantity that peaks at a level expected to diverge at the transition,
at a time that we may estimate as $t_{\alpha}$, see Fig \ref{chi4}.
All in all, we obtain the phase diagram of Fig \ref{phasediagram}.
We know that the $\tau^{-1}=0$ axis is just equivalent to the thermal problem,
with temperature $N/M$. What can we say about clustering for smaller
$\tau$? Figure \ref{concu1} shows the contributions of different
configurations (the top uniform color corresponds to contributions
smaller than 10\% each). We see that for all but the highest $\tau$,
the system is in the {\em concurrent mutation regime} \cite{desai_beneficial_2007},
and the thermal correspondence, applied naively, breaks down. Recalling
previous sections, we still expect that whenever the $\alpha$ relaxation
time is large (near and below the transition), the correspondence
with a thermal system may still hold, but taken for quantities that
are averaged over a state, and considering two situations at time-separations
larger than $t_{\alpha}$.

\vspace{.2cm}

{\bf A test of detailed balance in a region with large $\tau_\alpha$}

\vspace{.2cm}

Cheking detailed balance numerically is extremely hard. We shall here use the Fluctuation Theorem \cite{mustonen_fitness_2010,evans_fluctuation_2002}
derived above, as an indirect test. Because this theorem requires to start from equilibrium, we are only in a position to
do the test close to the glass transition, where the time-separation is large enough, but not within, because then equilibration
becomes problematic. 
  We thus place ourselves just {\em
above} the transition (so that a stationary distribution might be
reached) but not far from it (so that $t_{\alpha}$ is large), the
circled numbers in Figure \ref{phasediagram}. We start with a system
in equilibrium at time $t=0$, and we switch on a field $E\to E-hA$,
where $A$ is any observable, in our case we choose $A=\sum_{i}s_{i}$.
After an arbitrary time $t$ we measure again the value of $A$ and
check the equation (see \ref{ft-1}) 
\begin{equation}
\ln{P[A(t)-A(0)=\Delta]}-\ln{P[A(t)-A(0)=-\Delta]}=\beta h\Delta\label{ii}
\end{equation}

\begin{figure}
\centering \includegraphics[angle=270,width=0.48\columnwidth]{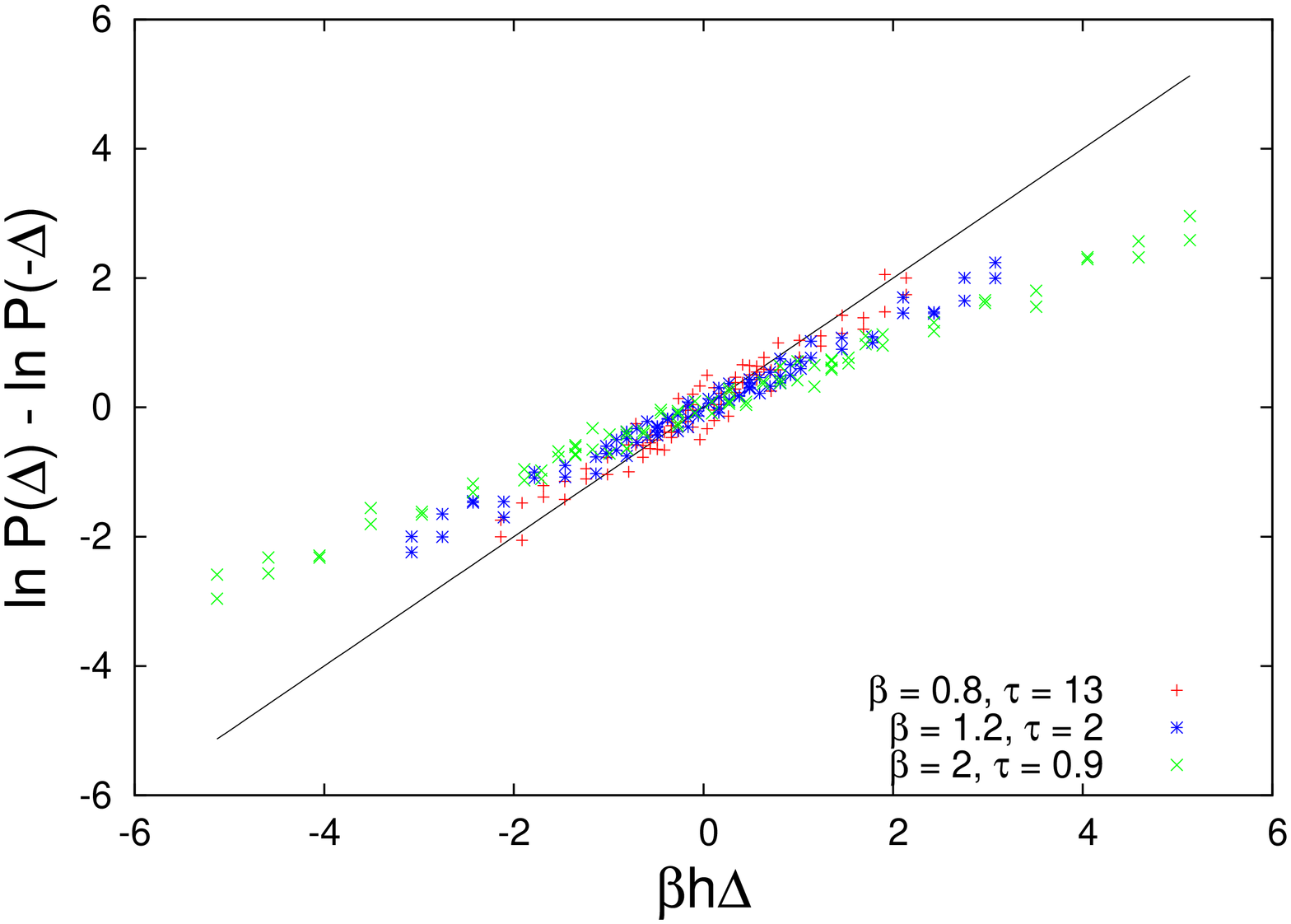}
\includegraphics[angle=270,width=0.48\columnwidth]{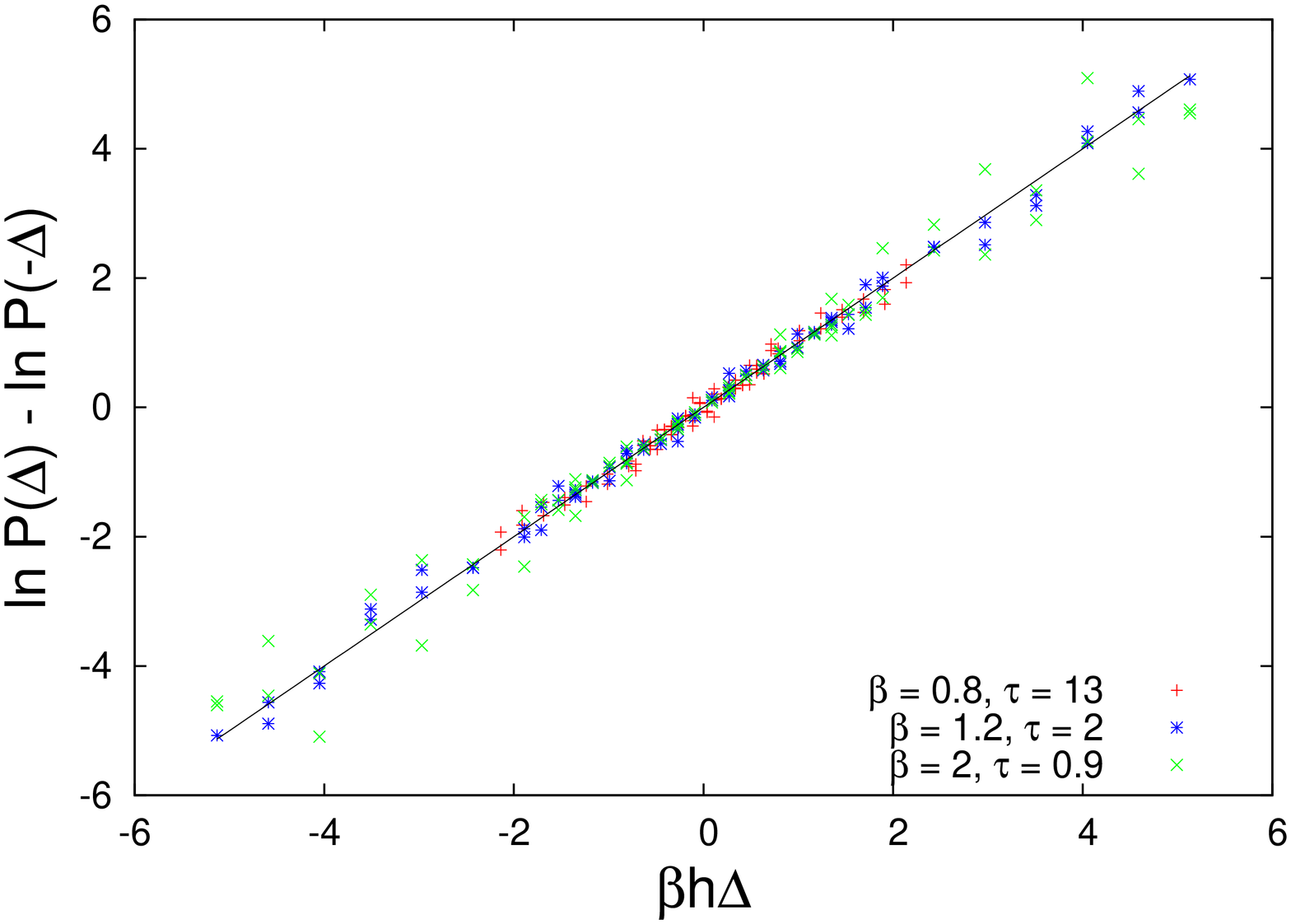}
\protect\protect\protect\protect\caption{$\ln P(\Delta)-\ln P(-\Delta)$ versus $\beta h\Delta$, for close times
and instantaneous values (left) and for quantities averaged over a
time-window, and times separated by more than $t_{\alpha}$ (right).
The latter works perfectly, without any fitting parameters. The parameters
for the three curves, $(\beta,\tau)={(0.8,13),(1.2,2),(2,0.9)}$, correspond
to those of figure \ref{concu1}, and are represented on the phase
diagram of figure \ref{phasediagram}.}
\label{FT} 
\end{figure}

We obtain the plot Fig \ref{FT} (left), which does not verify the
Fluctuation Theorem \ref{ii}, except for very large $\tau$. {
This is what we expected, as there is no clustering into a single
type, and the connection with a thermal system {\em fails}. Instead, when
we compute the differences as $\Delta=\bar{A}(t)-\bar{A}(t=0)$, with
$\bar{A}$ the average of $A$ within a window comparable to the time
to reach a plateau -- the {\em `equilibration within a super-state'} time $\ll t_{\alpha}$
-- and $t\sim3t_{\alpha}$, the relation (\ref{ii}) for the averaged
values $\Delta=\bar{A}(t)-\bar{A}(0)$ works perfectly, without fitting
parameters.} Note that below the transition line, the assumption
of equilibration fails (at least for a large system); because of this,
even if we expect detailed balance to hold for averaged quantities,
the Fluctuation Relation does not apply

Finally, note that in our XorSAT simulations we have made two choices. First, we work
with mutations that induce gradual jumps in the fitness (in sharp
contrast, for example, with the House-of-cards model). Secondly, we
chose the scale $N_{s}$ to be equal to the genome size $N$. As a
result, individual mutations incur a change in fitness $\delta\lambda_{mut}\sim1/M$.
The jumps of $\delta\lambda_{mut}$ are larger than would allow for
detailed balance to hold in a 1d step model, so a non-trivial extension
of the known limiting conditions has been demonstrated. The scaling
is the same as required by the step model, $\delta\lambda_{mut}\ll1/M$,
and therefore here we have not shown that the mutation steps can be
parametrically larger (as a function of $M$). The population width
in our simulations is $\sigma_{\lambda}\sim1/M$, narrower than the
minimum required for the step model: $\sigma_{\lambda}\sim M^{-1/2}$.

\begin{figure}
\centering \includegraphics[angle=270,width=0.48\columnwidth]{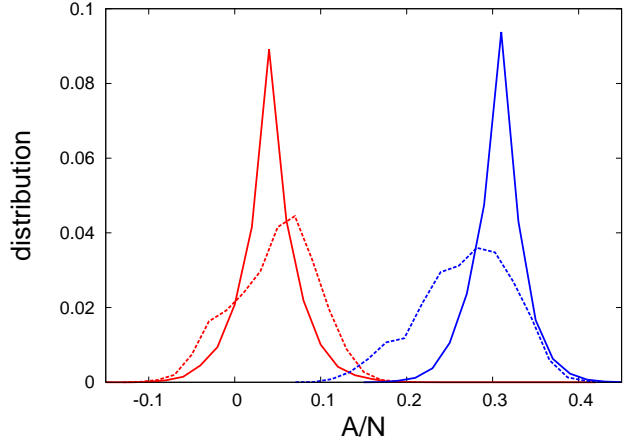}
\protect\protect\protect\protect\caption{Instantaneous distribution of $A$ for two times, and for two time-windows
of order $<t_{\alpha}$. The two sets are for two times separated
by $3t_{\alpha}$. Fluctuations inside a state and between states
are both larger than the instantaneous dispersion, underlying the
fact that the distribution within a state is not the $M=\infty$ one
predicted by the replicator equation.}
\label{cloud_width} 
\end{figure}

\section{Exploiting the correspondence }
\label{ExploitCorr}

Let us discuss a few examples where the correspondence between thermal
and population dynamics is exploited.

\vspace{0.5cm}

\textbf{Replica-exchange / Parallel Tempering}

\vspace{0.5cm}

\begin{figure}
\centering \includegraphics[width=7cm]{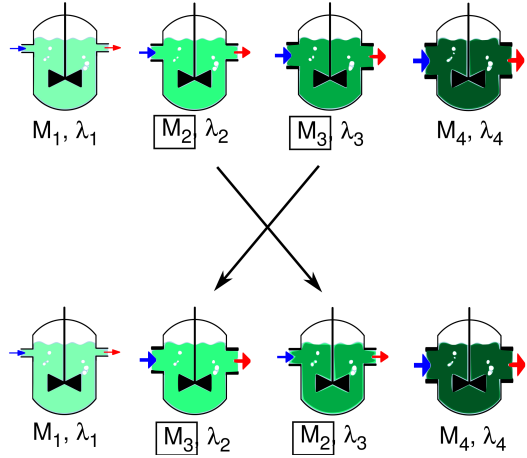} \protect\protect\protect\protect\caption{A series of chemostats with replica-exchange. \textcolor{black}{Different population sizes are obtained by regulating the flow of medium.}}
\label{tempering} 
\end{figure}

It is a well known fact that evolution speeds up when small groups
of individuals are in isolation from other groups \cite{hallatschek_genetic_2007,johnson_sewall_2008}.
In the language of this paper, this is in direct analogy with warming
up a complex system to make it anneal faster. Another related fact
is the facilitation in development of antibiotic resistance by concentration
gradients \cite{hermsen_rapidity_2012}. These two are extremely close
to being evolutionary counterparts of more sophisticated equilibration
procedures applied for complex systems.
Consider the configuration of Fig. \ref{tempering} 
The battery of chemostats are tuned to have increasing populations
$M_{1}<M_{2}<M_{3}...$ If we are able to measure the reproduction
rates of the cells, and exchange two chemostats with probability $\min\left\{ \left(\frac{\lambda_{j}}{\lambda_{i}}\right)^{M_{i}-M_{j}};1\right\} $
we obtain the procedure known as {\em Parallel Tempering} \cite{earl_parallel_2005}.
This is a well established procedure where several systems at different
temperatures are evolved separately, and occasionally permuted following
a rule analogous to the above. The interest is not only to accelerate
evolution, but to do so in a manner that \emph{respects the thermal
distributions}, in our case the distributions associated with each
population size. Thus the same final distribution is obtained, but
it is reached faster.

An alternative way of implementing the same procedure is the following.
Take two chemostats A and B, duplicate them into A' and B'. Seed an
individual of A' into B', and an individual of B' into bottle A'.
If {\em both} invading individuals are successful in fixing, replace
the (A,B) by (A',B'). Otherwise, discard (A',B') and continue with
(A,B). One can show that the probabilities are the correct ones for
Parallel Tempering, that is: 
\begin{equation}
\frac{P(i:A\to B\mathrm{\ \ and\ \ }j:B\to A)}{P(i:B\to A\mathrm{\ \ and\ \ }j:A\to B)}=e^{(\beta_{i}-\beta_{j})(E_{i}-E_{j})}
\end{equation}

Another version of Parallel Tempering is the `many fields', rather
than the `many temperature' version. The systems are then all at the
same temperature, but at different fields $h_{1}(S),h_{2}(S),...$.
Exchange between two systems is accepted with the Monte Carlo rule.
The analogue here is to use a series of chemostats with, for example,
increasing concentrations of antibiotic, and make exchanges in concentrations
according to the rule in the previous paragraph.

\vspace{0.5cm}

\textbf{Kovacs effect }

\vspace{0.5cm}

The Kovacs effect is a manifestation of the presence of processes
with more than one relaxation timescale within an out-of-equilibrium
system. The idea is to make the fast and slow processes `play against'
one another. Originally, it was discussed as a prove that a glass
cannot be represented by a single extra parameter encapsulating all
the history of temperature changes. To do this, Kovacs \cite{kovacs_transition_1964}
considered two out of equilibrium situations with equal energy but
with different previous histories, and showed that their subsequent
evolutions are different. In our present case, the Kovacs effect shows
that fitness need not increase, even when parameters are not changing.

In a glassy context \cite{mossa_crossover_2004}, the system is started
in the liquid phase and quenched rapidly to very low temperatures.
The glass so obtained evolves fast at first, but remains trapped in
regions of relatively high energy from which it finds it hard to escape.
Next, the system is taken to a higher, but still glassy, temperature.
The immediate reaction is an increase in energy, because of the thermal
agitation, but at later times the increased thermal activation allows
the system to optimize more efficiently, and the energy decreases.

\begin{figure}
\centering \includegraphics[angle=270,width=0.48\columnwidth]{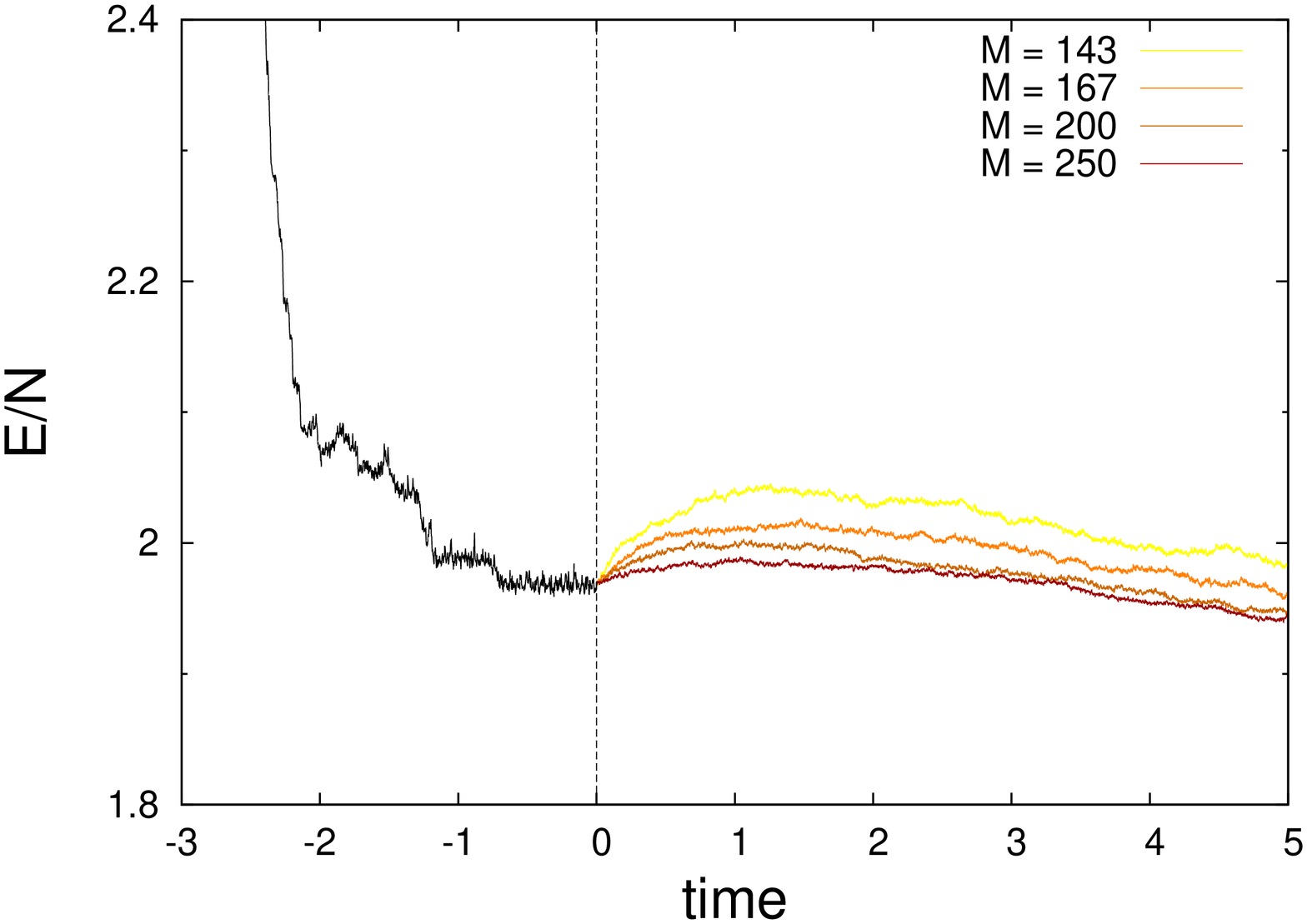}
\includegraphics[angle=270,width=0.48\columnwidth]{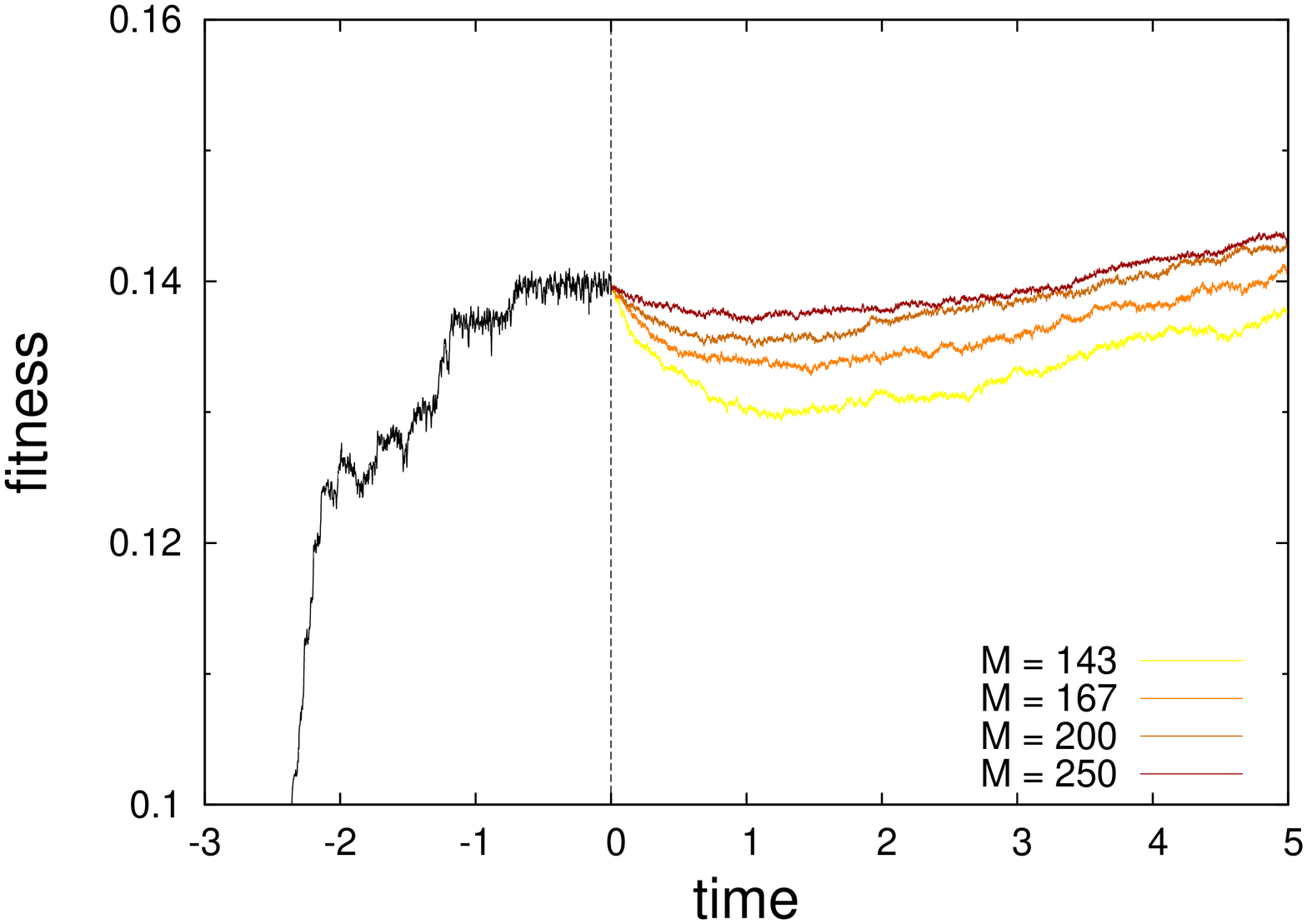}
\protect\protect\protect\protect\caption{Energy (left) and fitness (right) evolutions following adaptation
to a new situation, followed by a bottleneck. At some time $t<0$,
the population is quenched to a very low temperature, $M_{q}=1000$,
and let relax (black solid line). At time $t=0$ (vertical dashed
line) the population is reduced instantaneously to a lower M = \{143,
167, 200, 250\} (lighter to darker lines). The fast timescale re-adaptation
causes, in the short period, the energy (resp. fitness) to increase
(resp. decrease). However, on longer timescales, the smaller size
of the population allows it to explore more efficiently the energy
(fitness) landscape, finding more favourable configurations.}
\label{kovacs} 
\end{figure}

\begin{figure}
\centering \includegraphics[angle=270,width=0.48\columnwidth]{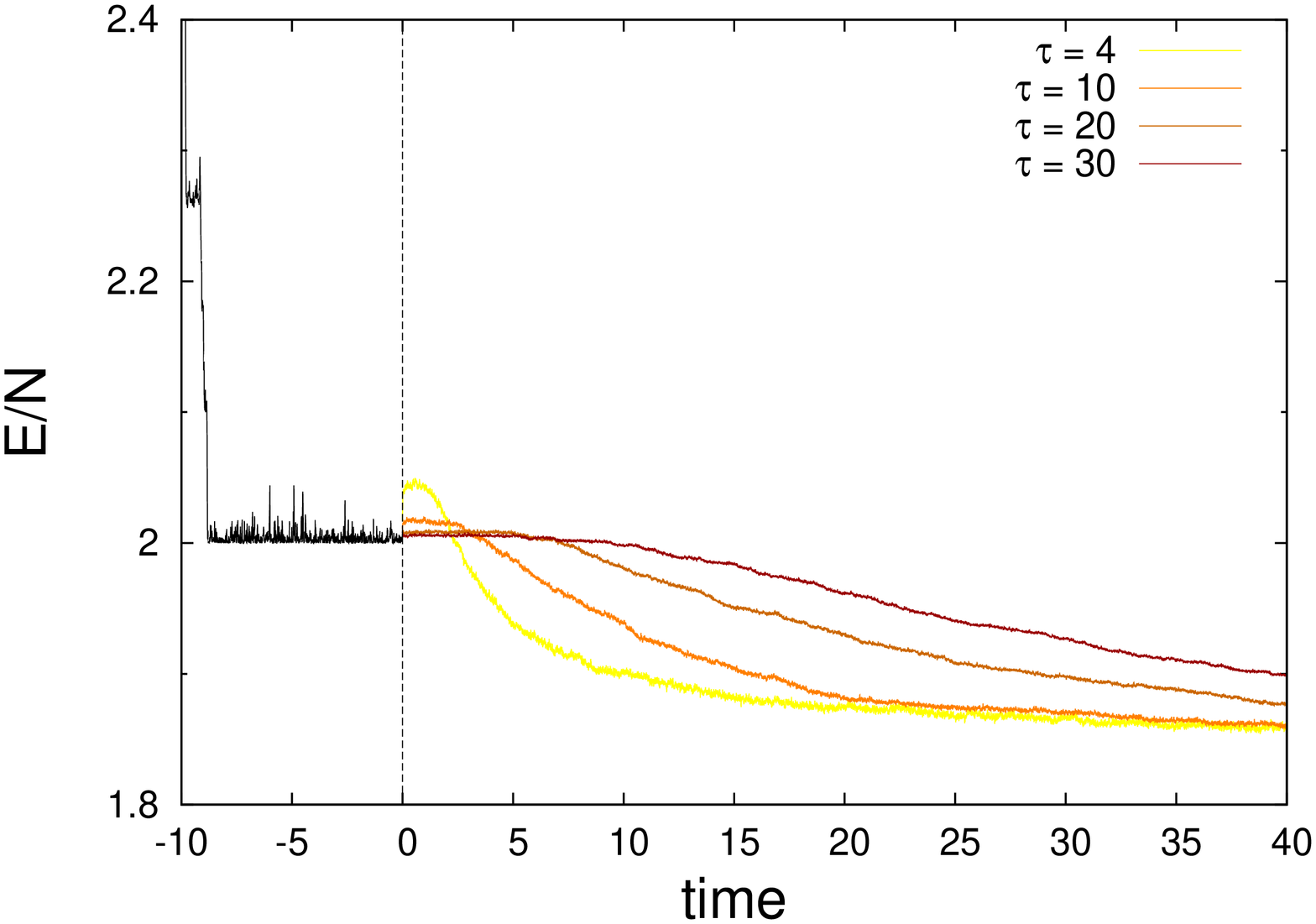}
\includegraphics[angle=270,width=0.48\columnwidth]{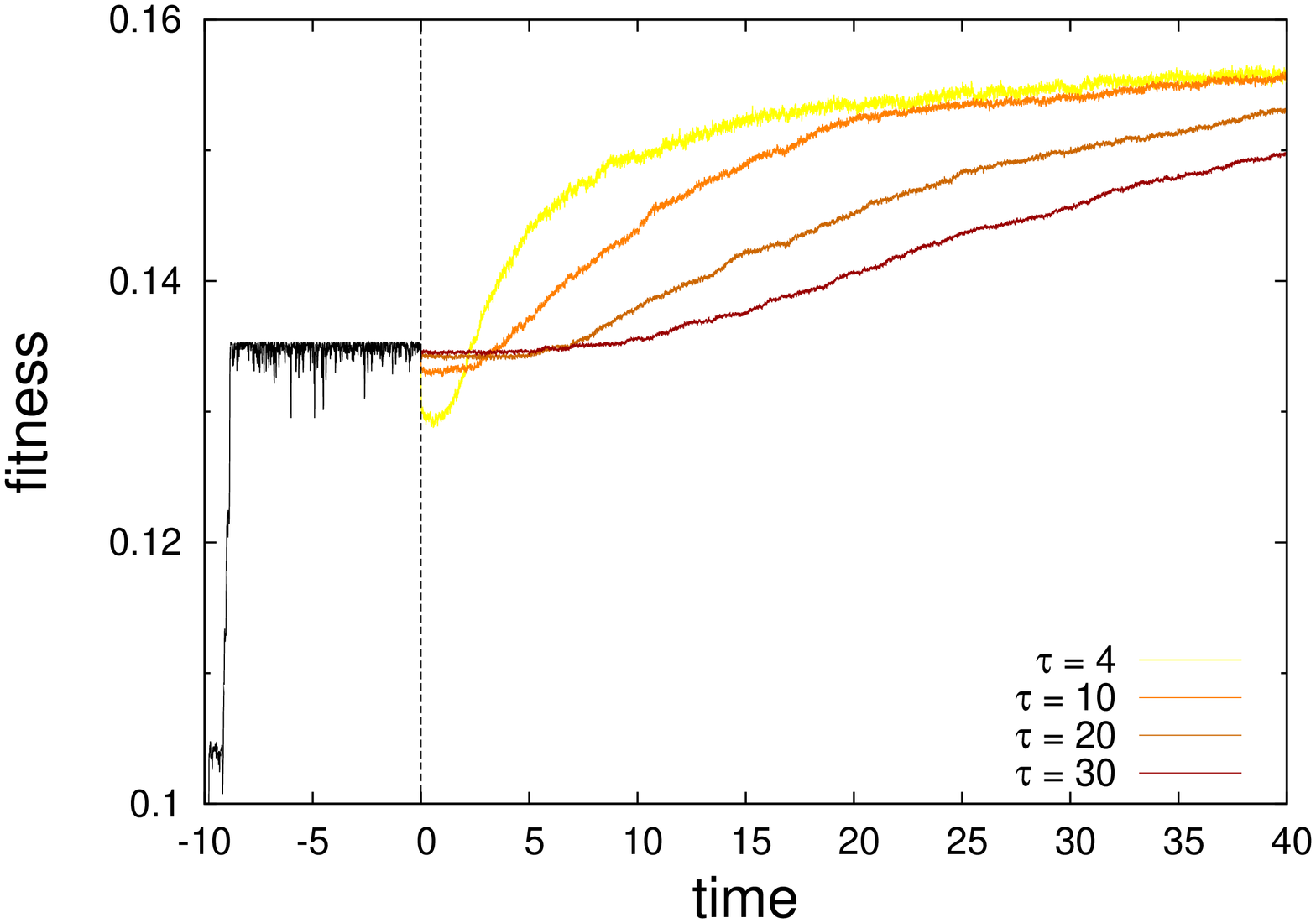}
\protect\protect\protect\protect\caption{Energy and fitness evolutions following an increase in mutation rate.
The population is let evolve at a low mutation rate, $\tau=40$ (black
solid line). At time $t=0$ (vertical dashed line) the mutation rate
is instantaneously increased, to value $\tau={4,10,20,30}$ (lighter
to darker lines). Again, the interplay between fast and slow timescale
causes the energy (resp. fitness) to increase (resp. decrease) at
first, to then decrease (increase) in the long run.}
\label{kovacstau} 
\end{figure}

For a population, the Kovacs effect would proceed as follows. A system
is put in a new, stressed situation. The population is allowed to
grow fast as it adapts. Once population has reached high levels, new
mutations become difficult to establish. If one then isolates a small
subsystem, and keeps its size small, at first some bad mutations will
decrease the fitness, but in the long run fitness should increase beyond
what it was originally, thanks to the possibility of a small system
to explore beneficial mutations. In Fig \ref{kovacs} we shows how
this happens in our simple model. Another variant of the same is,
instead of isolating a subensemble of the population, to consider
a situation in which the individuals suddenly increase their mutation
rate. The effect is similar, as shown in Fig. \ref{kovacstau}.


\vspace{0.5cm}

\textbf{More complex cycles}

\vspace{0.5cm}

\begin{figure}
\centering \includegraphics[angle=270,width=10cm]{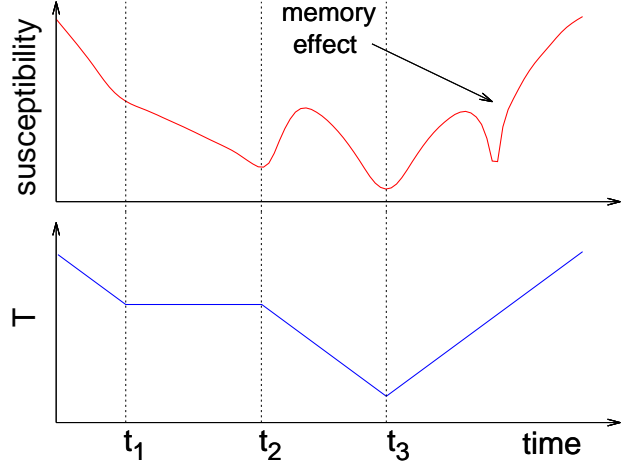} \protect\protect\protect\protect\caption{The experimental protocol of \cite{jonason_memory_1998}. \textcolor{black}{A procedure of annealing is carried on until $t=t_1$. Then, while $t_1<t<t_2$ the system is allowed to relax, while the temperature is keep fixed to $T_w$. The annealing is then resumed up to a time $t_3$. The system is subsequently subjected to a smooth increase in temperature: as the temperature passes through $T=T_w$ a drop in the susceptibility is observed, showing memory of the annealing that had taken place at that temperature.}}
\label{waiting} 
\end{figure}

Kovacs' effect is relatively simple. There are other effects in glassy
systems that are highly nontrivial, and have only been observed after
a very educated (by theory) guess stimulated the experiment with spin
glasses. The experimental protocol \cite{jonason_memory_1998} is
sketched in Fig \ref{waiting}. In the `writing period' the temperature
is slowly decreased starting from the high temperature phase down
to very low temperature, but stopping for relatively long times in
an intermediate temperature and waiting before resuming the decrease.
After reaching the lowest temperature, one starts the `reading' period,
which consists of a slow increase in temperature, while measuring
at the same time any quantity that may yield information of the `age'
of the sample, in the case of spin glasses the response to a low-frequency
a.c. field (which becomes smaller as time passes). The surprising
fact is that when the temperature during the `reading phase' traverses
the value at which there was a stop in the `writing phase', there
is a dip in the response to an a.c. field -- an indication of being
`older' {\em at that temperature}: as if the system could age independently
at different temperatures. This may be even generalized to several
stops, which are memorized independently. It is easy to propose a
similar experiment for a population, made to grow slowly, with stops
at different `bottlenecks'. Different subsamples of different sizes
extracted from the large population would bear memory of past history,
subsample sizes corresponding to  bottleneck sizes that happened in the past would show
measurable differences in their successful mutation rates (they should be smaller) with subsamples whose size does not correspond to  bottleneck sizes undergone in the past.

We have not attempted to simulate this here, because no successful
simulation has ever been performed yet for this model, even in the
original thermal context, probably due to size and time constraints
in the numerical power.

\vspace{0.5cm}

\textbf{Changing environments: connection to glassy rheology}

\vspace{0.5cm}

A system as the one we are considering, which is achieving better fitness
by slowly adapting to a complex landscape, is extremely sensitive
to changes in this landscape. This effect has been discussed in \cite{mustonen_molecular_2008},
although in a slightly different form, and also in  \cite{kussell_phenotypic_2005}. The counterpart in glass physics
of this fact has long been known. Consider the situation \cite{struik_rejuvenation_1997}
of a plastic bar prepared a time $t_{w}$ ago from a melt. The polymers
constituting the bar slowly rearrange -- ever more slowly -- to energetically
better and better configurations, and this process is known to go
on at least for decades. The bar is out of equilibrium, a fact that
we may recognize by testing its response to stress, which measurably
depends on $t_{w}$. Now suppose that we apply a large, fixed deformation
to the bar, for example applying a strong torsion one way and the other. The new constraints change the problem of optimization
the polymers are `solving': we expect evolution to restart to a certain
extent, and the apparent `age' of the bar to become smaller than $t_{w}$.
This is indeed what happens \cite{struik_rejuvenation_1997}, a phenomenon
called `rejuvenation'. Rejuvenation brings about an acceleration in
the dynamics. If the changes are continuous and different,  instead of aging (growth of $t_{\alpha}$), the system
settles in a value of $t_{\alpha}$ that depends on -- adapts to --
the speed of change of the energy landscape.  {\em Note that this property of evolution speed adapting
to landscape change speed, that is often attributed to a form of criticality \cite{kauffman_metabolic_1969}, here it appears as a universal
property of aging systems}.

\begin{figure}
\centering \includegraphics[width=6cm,angle=-90]{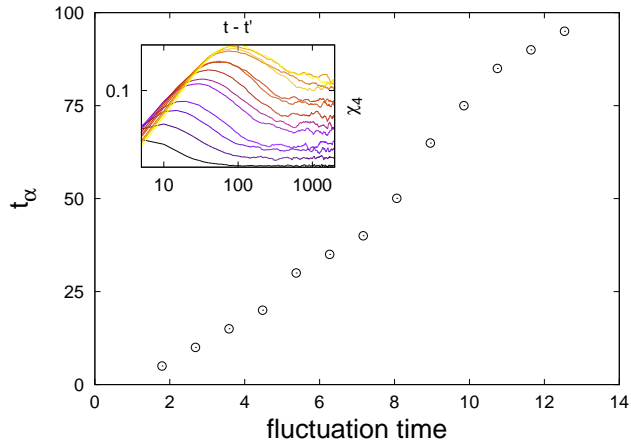} \protect\protect\protect\protect\caption{$\alpha$-time versus speed of variation of fitness landscape. Inset:
$\chi_{4}$ versus time for different speeds of random variation of landscape
(darker curves correspond to faster variations). The $t_{\alpha}$
value for each point of the main figure is the time at which the corresponding
curve reaches its maximum, a measure of the $\alpha$-time. }
\label{figa} 
\end{figure}

 Applying the same logic to our model, one expects a similar result. In order to model
 the changes in fitness landscape, we change at fixed intervals of time a randomly chosen clause, for example by changing the identity of one of the
 intervening Boolean variables (a slight change in the fitness function). For different rates of change, we plot the `age' of the system,
 as measured by $\chi_4$.  The results are shown on Figure
\ref{figa}: if the environment is randomly
changing, the system  evolves to accommodate various conditions,
and time-scales for changes in the environment are  reflected in
the time-scales for changes inside system.

Although we shall not pursue this line here, let us remark that one need not consider only random changes of fitness landscape, but
also repetitive ones. Recently, Fridman et al \cite{fridman_optimization_2014}
subjected bacterial populations to intermittent exposure to antibiotics.
All strains adapted via phenotypic changes and developed of tolerance
by adjusting the lag time of bacteria before regrowth to match the
duration of the antibiotic-exposure interval. In other words, the
system adapted to the cycle itself. Correspondingly, in another recent
paper, Fiocco\emph{ et. al.} \cite{fiocco_encoding_2014} have studied
the effect of letting evolve a glassy system under the influence of
a periodic field, strong enough to affect substantially its evolution.
The system `adapts' to this non-stationary situation, just as it would
to a stationary field: subjecting further the sample to new cycles
of different amplitudes (`reading') one may easily distinguish the
cycle at which it has been optimized (`writing'). 

\section{Discussion }
\label{Discussion}

Let us discuss here the domain of validity of the mapping into a thermal
system. The first element we have assumed throughout, is the fact
that the population is composed of individuals that do not interact,
except through a mechanism that limits the total population size.
More general settings could include the situation where the limited
resource is space, as in the case of cells growing in a Petri dish.
It is not impossible that generalizations in this direction can be
made.

We have lifted the assumption of successional mutations. The price
we pay is the fact that the thermal approach is then only valid for
the slow evolution; for the fast mutations, the detailed balance property
does not hold. The advantage is that now the approach is valid to
any system evolving through random drift and selection if the fitness
landscape is complex enough, so that the optimum is not reached in
reasonable times, and evolution slows down.

The most general  conditions for our extended detailed
balance to hold are not yet known, and will require further investigation.
Mutation probabilities (the $\mu_{ij}$) need to satisfy a symmetry
property, Eq. (\ref{detba}), for detailed balance to hold. If, on
the contrary, the mutation dynamics at the level of the individuals
contains cycles, the detailed balance property will be violated. Two
things may happen: either the cycles persist even for the slow timescales
(so that there are closed cycles of slow mutations), or they do not.
In the latter case slow dynamics would still be thermal.

As was mentioned in several places, we must give up the property of
equilibrium, which is not realistic even when detailed balance holds.
We are left with a system that is in fact like a glass: it works its
way towards equilibrium while external conditions do not change, but
never achieves it. Such systems have their own properties and rules.
We have now a much better understanding of them, and there is the
potential that many of the features that we have encountered in studying
glasses may be applied to evolving populations with strong epistasis.

\acknowledgments We would like to thank JP Bouchaud, and D.A. Kessler
for helpful discussions.

\bibliographystyle{IEEEtran}
\bibliography{IEEEabrv,paper}

\end{document}